\documentclass[12pt]{article}

\usepackage{amstex}

\begin{document}
\vspace*{-.6in}
\thispagestyle{empty}
\begin{flushright}
CALT-68-2293\\
CITUSC/00-045\\
hep-ex/0008017
\end{flushright}
\baselineskip = 18pt

\vspace{.5in}
{\Large
\begin{center}
Introduction to Superstring Theory
\end{center}}

\begin{center}
John H. Schwarz\footnote{Work
supported in part by the U.S. Dept. of Energy under Grant No.
DE-FG03-92-ER40701.}\\
\emph{California Institute of Technology\\ Pasadena, CA  91125, USA}
\end{center}
\vspace{1in}

\begin{center}
\textbf{Abstract}
\end{center}
\begin{quotation}
\noindent
These four lectures, addressed to an audience of graduate students in
experimental high energy physics, survey some of the basic concepts in string
theory.  The purpose is to convey a general sense of what string theory is and
what it has achieved. Since the characteristic scale of string theory is expected to
be close to the Planck scale, the structure of strings probably cannot be probed
directly in accelerator experiments.
The most accessible experimental implication of superstring theory
is supersymmetry at or below the TeV scale.
\end{quotation}

\vfil
\centerline{\it Lectures presented at the NATO Advanced Study Institute}
\centerline{\it on Techniques and Concepts of High Energy Physics}
\centerline{\it St. Croix, Virgin Islands --- June 2000}

\newpage

\pagenumbering{arabic}

\tableofcontents

\newpage

\section*{Introduction}

Tom Ferbel has presented me with a large challenge: explain string
theory to an audience of graduate students in experimental high
energy physics.  The allotted time is four 75-minute lectures.
This should be possible, if the goals are realistic. One goal is
to give a general sense of what the subject is about, and why
so many theoretical physicists are enthusiastic about it. Perhaps
you should regard these lectures as a cultural experience
providing a window into the world of abstract theoretical physics.
Don't worry if you miss some of the technical details in the
second and third lectures.  There is only one message in these
lectures that is important for experimental research: low-energy
supersymmetry is very well motivated theoretically, and it
warrants the intense effort that is being made to devise ways of
observing it. There are other facts that are nice to know,
however. For example, consistency of quantum theory and gravity is
a severe restriction, with farreaching consequences.

As will be explained, string theory requires supersymmetry, and
therefore string theorists were among the first to discover it.
Supersymmetric string theories are called superstring theories. At
one time there seemed to be five distinct superstring theories,
but it was eventually realized that each of them is actually a
special limiting case of a completely unique underlying theory.
This theory is not yet fully formulated, and when it is, we might
decide that a new name is appropriate.  Be that as it may, it is
clear that we are exploring an extraordinarily rich structure with
many deep connections to various branches of fundamental
mathematics and theoretical physics.  Whatever the ultimate status
of this theory may be, it is clear that these studies have already
been a richly rewarding experience.

To fully appreciate the mathematical edifice underlying
superstring theory requires an investment of time and effort. Many
theorists who make this investment really become hooked by it, and
then there is no turning back. Well, hooking you in this way is
not my goal, since you are engaged in other important activities;
but hopefully these lectures will convey an idea of why many
theorists find the subject so enticing. For those who wish to
study the subject in more detail, there are two standard textbook
presentations \cite{GSW, JPbook}.

The plan of these lectures is as follows:  The first lecture will
consist of a general non-technical overview of the subject.  It is
essentially the current version of my physics colloquium lecture.
It will describe some of the basic concepts and issues without
technical details.  If successful, it will get you sufficiently
interested in the subject that you are willing to sit through some
of the basic nitty-gritty analysis that explains what we mean by a
relativistic string, and how its normal modes are analyzed.
Lecture 2 will present the analysis for the bosonic string theory.
This is an unrealistic theory, with bosons only, but its study is
a pedagogically useful first step.  It involves many, but not all,
of the issues that arise for superstrings.  In lecture 3 the
extension to incorporate fermions and supersymmetry is described.
There are two basic formalisms for doing this (called RNS and GS).
Due to time limitations, only the first of these will be presented
here.  The final lecture will survey some of the more recent
developments in the field.  These include various nonperturbative
dualities, the existence of an 11-dimensional limit (called
M-theory) and the existence of extended objects of various
dimensionalities, called $p$-branes.  As will be explained, a
particular class of $p$-branes, called D-branes, plays an
especially important role in modern research.

\section{Lecture 1:  Overview and Motivation}

Many of the major developments in fundamental physics of the past
century arose from identifying and overcoming contradictions
between existing ideas.  For example, the incompatibility of
Maxwell's equations and Galilean invariance led Einstein to
propose the special theory of relativity.  Similarly, the
inconsistency of special relativity with Newtonian gravity led him
to develop the general theory of relativity.  More recently, the
reconciliation of special relativity with quantum mechanics led to
the development of quantum field theory.  We are now facing
another crisis of the same character.  Namely, general relativity
appears to be incompatible with quantum field theory.  Any
straightforward attempt to ``quantize'' general relativity leads
to a nonrenormalizable theory.  In my opinion, this means that the
theory is inconsistent and needs to be modified at short distances
or high energies.  The way that string theory does this is to give
up one of the basic assumptions of quantum field theory, the
assumption that elementary particles are mathematical points, and
instead to develop a quantum field theory of one-dimensional
extended objects, called strings,  There are very few consistent
theories of this type, but superstring theory shows great promise
as a unified quantum theory of all fundamental forces including
gravity.  There is no realistic string theory of elementary
particles that could serve as a new standard model, since there is
much that is not yet understood. But that, together with a deeper
understanding of cosmology, is the goal. This is still a work in
progress.

Even though string theory is not yet fully formulated, and we cannot yet give a
detailed description of how the standard model of elementary particles should
emerge at low energies, there are some general features of the theory that can
be identified.  These are features that seem to be quite generic irrespective
of how various details are resolved.  The first, and perhaps most important, is
that general relativity is necessarily incorporated in the theory.  It gets
modified at very short distances/high energies but at ordinary distance and
energies it is present in exactly the form proposed by Einstein.  This is
significant, because it is arising within the framework of a consistent quantum
theory.  Ordinary quantum field theory does not allow gravity to exist; string
theory requires it!  The second general fact is that Yang--Mills gauge theories
of the sort that comprise the standard model naturally arise in string theory.
We do not understand why the specific $SU(3) \times SU(2) \times U(1)$ gauge
theory of the standard model should be preferred, but (anomaly-free) theories
of this general type do arise naturally at ordinary energies.  The third
general feature of string theory solutions is supersymmetry.  The mathematical
consistency of string theory depends crucially on supersymmetry, and it is very
hard to find consistent solutions (quantum vacua) that do not preserve at least
a portion of this supersymmetry.  This prediction of string theory differs from
the other two (general relativity and gauge theories) in that it really is a
prediction.  It is a generic feature of string theory that has not yet been
discovered experimentally.

\subsection{Supersymmetry}

Even though supersymmetry is a very important part of the story,
the discussion here will be very brief, since it will be discussed
in detail by other lecturers. There will only be a few general
remarks. First, as we have just said, supersymmetry is the major
prediction of string theory that could appear at accessible
energies, that has not yet been discovered. A variety of
arguments, not specific to string theory, suggest that the
characteristic energy scale associated to supersymmetry breaking
should be related to the electroweak scale, in other words in the
range 100 GeV -- 1 TeV. The symmetry implies that all known
elementary particles should have partner particles, whose masses
are in this general range. This means that some of these
superpartners should be observable at the CERN Large Hadron
Collider (LHC), which will begin operating in the middle part of
this decade.  There is even a chance that Fermilab Tevatron
experiments could find superparticles earlier than that.

In most versions of phenomenological supersymmetry there is a
multiplicatively conserved quantum number called R-parity. All
known particles have even R-parity, whereas their superpartners
have odd R-parity. This implies that the superparticles must be
pair-produced in particle collisions.  It also implies that the
lightest supersymmetry particle (or LSP) should be absolutely
stable. It is not known with certainty which particle is the LSP,
but one popular guess is that it is a ``neutralino.'' This is an
electrically neutral fermion that is a quantum-mechanical mixture
of the partners of the photon, $Z^0$, and neutral Higgs particles.
Such an LSP would interact very weakly, more or less like a
neutrino.  It is of considerable interest,  since it is an
excellent dark matter candidate. Searches for dark matter
particles called WIMPS (weakly interacting massive particles)
could discover the LSP some day. Current experiments might not
have sufficient detector volume to compensate for the exceedingly
small cross sections.

There are three unrelated arguments that point to the same mass
range for superparticles.  The one we have just been discussing, a
neutralino LSP as an important component of dark matter, requires
a mass of order 100 GeV.  The precise number depends on the
mixture that comprises the LSP, what their density is, and a
number of other details. A second argument is based on the famous
hierarchy problem. This is the fact that standard model radiative
corrections tend to renormalize the Higgs mass to a very high
scale.  The way to prevent this is to extend the standard model to
a supersymmetric standard model and to have the supersymmetry be
broken at a scale comparable to the Higgs mass, and hence to the
electroweak scale. The third argument that gives an estimate of
the susy-breaking scale is grand unification. If one accepts the
notion that the standard model gauge group is embedded in a larger
gauge group such as $SU(5)$ or $SO(10)$, which is broken at a high
mass scale, then the three standard model coupling constants
should unify at that mass scale.  Given the spectrum of particles,
one can compute the evolution of the couplings as a function of
energy using renormalization group equations.  One finds that if
one only includes the standard model particles this unification
fails quite badly.  However, if one also includes all the
supersymmetry particles required by the minimal supersymmetric
extension of the standard model, then the couplings do unify at an
energy of about $2 \times 10^{16}$ GeV.  For this agreement to
take place, it is necessary that the masses of the superparticles
are less than a few TeV.

There is other support for this picture, such as the ease with
which supersymmetric grand unification explains the masses of the
top and bottom quarks and electroweak symmetry breaking. Despite
all these indications, we cannot be certain that this picture is
correct until it is demonstrated experimentally.  One could
suppose that all this is a giant coincidence, and the correct
description of TeV scale physics is based on something entirely
different. The only way we can decide for sure is by doing the
experiments. As I once told a newspaper reporter, in order to be
sure to be quoted: discovery of supersymmetry would be more
profound than life on Mars.

\subsection{Basic Ideas of String Theory}

In conventional quantum field theory the elementary particles are
mathematical points, whereas in perturbative string theory the
fundamental objects are one-dimensional loops (of zero thickness).
Strings have a characteristic length scale, which can be estimated
by dimensional analysis.  Since string theory is a relativistic
quantum theory that includes gravity it must involve the
fundamental constants $c$ (the speed of light), $\hbar$ (Planck's
constant divided by $2\pi$), and $G$ (Newton's gravitational
constant). From these one can form a length, known as the Planck
length
\begin{equation}
\ell_p = \left(\frac{\hbar G}{c^3}\right)^{3/2} = 1.6 \times 10^{-33} \, {\rm cm.}
\end{equation}
Similarly, the Planck mass is
\begin{equation}
m_p = \left(\frac{\hbar c}{G}\right)^{1/2} = 1.2 \times 10^{19} \, {\rm GeV}/c^2.
\end{equation}
Experiments at energies far below the Planck energy cannot resolve distances as
short as the Planck length.  Thus, at such energies, strings can be
accurately approximated by point particles.  From the viewpoint of string
theory, this explains why quantum field theory has been so successful.

As a string evolves in time it sweeps out a two-dimensional
surface in spacetime, which is called the world sheet of the
string.  This is the string counterpart of the world line for a
point particle.  In quantum field theory, analyzed in perturbation
theory, contributions to amplitudes are associated to Feynman
diagrams, which depict possible configurations of world lines.  In
particular, interactions correspond to junctions of world lines.
Similarly, string theory perturbation theory involves string world
sheets of various topologies.  A particularly significant fact is
that these world sheets are generically smooth.  The existence of
interaction is a consequence of world-sheet topology rather than a
local singularity on the world sheet.  This difference from
point-particle theories has two important implications.  First, in
string theory the structure of interactions is uniquely determined
by the free theory.
 There are no arbitrary interactions to be chosen.  Second, the ultraviolet
divergences of point-particle theories can be traced to the fact that
interactions are associated to world-line junctions at specific spacetime
points.  Because the string world sheet is smooth, string theory amplitudes
have no ultraviolet divergences.

\subsection{A Brief History of String Theory}

String theory arose in the late 1960's out of an attempt to describe the strong
nuclear force.  The inclusion of fermions led to the discovery of
supersymmetric strings --- or superstrings --- in 1971.  The subject fell out
of favor around 1973 with the development of QCD, which was quickly recognized
to be the correct theory of strong interactions.  Also, string theories had
various unrealistic features such as extra dimensions and massless particles,
neither of which are appropriate for a hadron theory.

Among the massless string states there is one that has spin two.
In 1974, it was shown by Scherk and me \cite{Scherk74}, and
independently by Yoneya \cite{Yoneya74}, that this particle
interacts like a graviton, so the theory actually includes general
relativity. This led us to propose that string theory should be
used for unification rather than for hadrons.  This implied, in
particular, that the string length scale should be comparable to
the Planck length, rather than the size of hadrons ($10^{-13}$ cm)
as we had previously assumed.

In the ``first superstring revolution,'' which took place in
1984--85, there were a number of important developments (described
later) that convinced a large segment of the theoretical physics
community that this is a worthy area of research.  By the time the
dust settled in 1985 we had learned that there are five distinct
consistent string theories, and that each of them requires
spacetime supersymmetry in the ten dimensions (nine spatial
dimensions plus time).  The theories, which will be described
later, are called type I, type IIA, type IIB, $SO(32)$ heterotic,
and $E_8 \times E_8$ heterotic.

\subsection{Compactification}

In the context of the original goal of string theory -- to explain
hadron physics -- extra dimensions are unacceptable. However, in a
theory that incorporates general relativity, the geometry of
spacetime is determined dynamically. Thus one could imagine that
the theory admits consistent quantum solutions in which the six
extra spatial dimensions form a compact space, too small to have
been observed. The natural first guess is that the size of this
space should be comparable to the string scale and the Planck
length.  Since the equations must be satisfied, the geometry of
this six-dimensional space is not arbitrary.  A particularly
appealing possibility, which is consistent with the equations, is
that it forms a type of space called a Calabi--Yau space
\cite{Candelas}.

Calabi--Yau compactification, in the context of the
$E_8 \times E_8$ heterotic string theory, can give a low-energy effective
theory that closely resembles a supersymmetric extension of the standard model.
There is actually a lot of freedom, because there are very many different
Calabi--Yau spaces, and there are other arbitrary choices that can be made.
Still, it is interesting that one can come quite close to realistic physics.
It is also interesting that the number of quark and lepton families that one
obtains is determined by the topology of the Calabi--Yau space.  Thus, for
suitable choices, one can arrange to end up with exactly three families.  People
were very excited by the picture in 1985.  Nowadays, we tend to make a more
sober appraisal that emphasizes all the arbitrariness that is involved, and the
things that don't work exactly right.  Still, it would not be surprising if
some aspects of this picture survive as part of the story when we understand
the right way to describe the real world.

\subsection{Perturbation Theory}

Until 1995 it was only understood how to formulate string theories
in terms of perturbation expansions.  Perturbation theory is
useful in a quantum theory that has a small dimensionless coupling
constant, such as quantum electrodynamics, since it allows one to
compute physical quantities as power series expansions in the
small parameter. In QED the small parameter is the fine-structure
constant $\alpha \sim 1/137$. Since this is quite small,
perturbation theory works very well for QED.  For a physical
quantity $T(\alpha)$, one computes (using Feynman diagrams)
\begin{equation}
T(\alpha) = T_0 + \alpha T_1 + \alpha^2 T_2 + \ldots.
\end{equation}
It is the case generically in quantum field theory that expansions of this type
are divergent.  More specifically, they are asymptotic expansions with zero
radius convergence.  Nonetheless, they can be numerically useful if the
expansion parameter is small.  The problem is that there are various
non-perturbative contributions (such as instantons) that have the structure
\begin{equation}
T_{NP} \sim e^{-(const./\alpha)}.
\end{equation}
In a theory such as QCD, there are regimes where perturbation theory is useful
(due to asymptotic freedom) and other regimes where it is not.  For problems of
the latter type, such as computing the hadron spectrum, nonperturbative methods
of computation, such as lattice gauge theory, are required.

In the case of string theory the dimensionless string coupling constant,
denoted $g_s$, is
determined dynamically by the expectation value of a scalar field called the
dilaton.  There is no particular reason that this number should be small.  So
it is unlikely that a realistic vacuum could be analyzed accurately using
perturbation theory.  More importantly, these theories have many qualitative
properties that are inherently nonperturbative.  So one needs nonperturbative
methods to understand them.

\subsection{The Second Superstring Revolution}

Around 1995 some amazing and unexpected ``dualities'' were discovered that
provided the first glimpses into nonperturbative features of string theory.
These dualities were quickly recognized to have three major implications.

The dualities enabled us to relate all five of the superstring
theories to one another.  This meant that, in a fundamental sense,
they are all equivalent to one another.  Another way of saying
this is that there is a unique underlying theory, and what we had
been calling five theories are better viewed as perturbation
expansions of this underlying theory about five different points
(in the space of consistent quantum vacua).  This was a profoundly
satisfying realization, since we really didn't want five theories
of nature.  That there is a completely unique theory, without any
dimensionless parameters, is the best outcome one could have hoped
for.  To avoid confusion, it should be emphasized that even though
the theory is unique, it is entirely possible that there are many
consistent quantum vacua.  Classically, the corresponding
statement is that a unique equation can admit many solutions.  It
is a particular solution (or quantum vacuum) that ultimately must
describe nature. At least, this is how a particle physicist would
say it.  If we hope to understand the origin and evolution of the
universe, in addition to  properties of elementary particles, it
would be nice if we could also understand cosmological solutions.

A second crucial discovery was that the theory admits a variety of
nonperturbative excitations, called $p$-branes, in addition to the fundamental
strings.  The letter $p$ labels the number of spatial dimensions of the
excitation.  Thus, in this language, a point particle is a 0-brane, a string is
a 1-brane, and so forth.  The reason that $p$-branes were not discovered in
perturbation theory is that they have tension (or energy density) that diverges
as $g_s \rightarrow 0$.  Thus they are absent from the perturbative theory.

The third major discovery was that the underlying theory also has
an eleven-dimensional solution, which is called M-theory.  Later,
we will explain how the eleventh dimension arises.

One type of duality is called S duality.  (The choice of the letter S is a
historical accident of no great significance.)
Two string theories (let's call them A and B) are
related by S duality if one of them evaluated at strong coupling is
equivalent to the other one evaluated at weak coupling.  Specifically, for any
physical quantity $f$, one has
\begin{equation}
f_A (g_s) = f_B (1/g_s).
\end{equation}
Two of the superstring theories --- type I and $SO(32)$ heterotic
--- are related by S duality in this way.  The type IIB theory is
self-dual.  Thus S duality is a symmetry of the IIB theory, and
this symmetry is unbroken if $g_s = 1$. Thanks to S duality, the
strong-coupling behavior of each of these three theories is
determined by a weak-coupling analysis. The remaining two
theories, type IIA and $E_8 \times E_8$ heterotic, behave very
differently at strong coupling.  They grow an eleventh dimension!

Another astonishing duality, which goes by the name of T duality, was
discovered several years earlier.  It can be understood in perturbation theory,
which is why it was found first.  But, fortunately, it often continues to be
valid even at strong coupling.  T duality can relate different
compactifications of different theories.  For example, suppose theory $A$ has a
compact dimension that is a circle of radius $R_A$ and theory $B$ has a compact
dimension that is a circle of radius $R_B$.  If these two theories are related
by T duality this means that they are equivalent provided that
\begin{equation}
R_A R_B = (\ell_s)^2,
\end{equation}
where $\ell_s$ is the fundamental string length scale. This has
the amazing implication that when one of the circles becomes small
the other one becomes large.  In a later lecture, we will explain
how this is possible.  T duality relates the two type II theories
and the two heterotic theories.  There are more complicated
examples of the same phenomenon involving compact spaces that are
more complicated than a circle, such as tori, K3, Calabi--Yau
spaces, etc.

\subsection{The Origins of Gauge Symmetry}

There are a variety of mechanisms than can give rise to
Yang--Mills type gauge symmetries in string theory.  Here, we will
focus on two basic possibilities: Kaluza--Klein symmetries and
brane symmetries.

The basic Kaluza--Klein idea goes back to the 1920's, though it
has been much generalized since then.  The idea is to suppose that
the 10- or 11-dimensional geometry has a product structure $M
\times K$, where $M$ is Minkowski spacetime and $K$ is a compact
manifold.  Then, if $K$ has symmetries, these appear as gauge
symmetries of the effective theory defined on $M$.  The
Yang--Mills gauge fields arise as components of the gravitational
metric field with one direction along $K$ and the other along $M$.
For example, if the space $K$ is an $n$-dimensional sphere, the
symmetry group is $SO(n+1)$, if it is $CP^n$ --- which has $2n$
dimensions --- it is $SU(n+1)$, and so forth.  Elegant as this may
be, it seems unlikely that a realistic $K$ has any such
symmetries. Calabi--Yau spaces, for example, do not have any.

A rather more promising way of achieving realistic gauge
symmetries is via the brane approach.  Here the idea is that a
certain class of $p$-branes (called D-branes) have gauge fields
that are restricted to their world volume.  This means that the
gauge fields are not defined throughout the 10- or 11-dimensional
spacetime but only on the $(p+1)$-dimensional hypersurface defined
by the D-branes.  This picture suggests that the world we observe
might be a D-brane embedded in a higher-dimensional space.  In
such a scenario, there can be two kinds of extra dimensions:
compact dimensions along the brane and compact dimensions
perpendicular to the brane.

The traditional viewpoint, which in my opinion is still the best
bet, is that all extra dimensions (of both types) have sizes of
order $10^{-30}$ to $10^{-32}$ cm corresponding to an energy scale
of $10^{16} - 10^{18}$ GeV.  This makes them inaccessible to
direct observation, though their existence would have definite
low-energy consequences.  However, one can and should ask ``what
are the experimental limits?''  For compact dimensions along the
brane, which support gauge fields, the nonobservation of extra
dimensions in tests of the standard model implies a bound of about
1 TeV. The LHC should extend this to about 10 TeV.  For compact
dimensions ``perpendicular to the brane,'' which only support
excitations with gravitational strength forces, the best bounds
come from Cavendish-type experiments, which test the $1/R^2$
structure of the Newton force law at short distances.  No
deviations have been observed to a distance of about 1 mm, so far.
Experiments planned in the near future should extend the limit to
about 100 microns. Obviously, observation of any deviation from
$1/R^2$ would be a major discovery.

\subsection{Conclusion}

This introductory lecture has sketched some of the remarkable
successes that string theory has achieved over the past 30 years.
There are many others that did not fit in this brief survey.
Despite all this progress, there are some very important and
fundamental questions whose answers are unknown. It seems that
whenever a breakthrough occurs, a host of new questions arise, and
the ultimate goal still seems a long way off. To convince you that
there is a long way to go, let us list some of the most important
questions:

\begin{itemize}
\item What is the theory?  Even though a great deal is known about string
theory and M theory, it seems that the optimal formulation of the underlying
theory has not yet been found.  It might be based on principles that have not
yet been formulated.

\item We are convinced that supersymmetry is present at high energies and
probably at the electroweak scale, too.  But we do not know how or why it is
broken.

\item A very crucial problem concerns the energy density of the vacuum, which
is a physical quantity in a gravitational theory.  This is
characterized by the cosmological constant, which observationally
appears to have a small positive value --- so that the vacuum
energy of the universe is comparable to the energy in matter.  In
Planck units this is a tiny number $(\Lambda \sim 10^{-120})$. If
supersymmetry were unbroken, we could argue that $\Lambda = 0$,
but if it is broken at the 1 TeV scale, that would seem to suggest
$\Lambda \sim 10^{-60}$, which is very far from the truth. Despite
an enormous amount of effort and ingenuity, it is not yet clear
how superstring theory will conspire to break supersymmetry at the
TeV scale and still give a value for $\Lambda$ that is much
smaller than $10^{-60}$. The fact that the desired result is about
the square of this might be a useful hint.

\item Even though the underlying theory is unique, there seem to
be many consistent quantum vacua.  We would very much like to
formulate a theoretical principle (not based on observation) for
choosing among these vacua.  It is not known whether the right
approach to the answer is cosmological, probabilistic, anthropic,
or something else.
\end{itemize}

\section{Lecture 2:  String Theory Basics}

In this lecture we will describe the world-sheet dynamics of the
original bosonic string theory.  As we will see this theory has
various unrealistic and unsatisfactory properties.  Nonetheless it
is a useful preliminary before describing supersymmetric strings,
because it allows us to introduce many of the key concepts without
simultaneously addressing the added complications associated with
fermions and supersymmetry.

We will describe string dynamics from a first-quantized point of
view.  This means that we focus on understanding it from a
world-sheet sum-over-histories point of view.  This approach is
closely tied to perturbation theory analysis. It should be
contrasted with ``second quantized'' string field theory which is
based on field operators that create or destroy entire strings.
Since the first-quantized point of view may be less familiar to
you than second-quantized field theory, let us begin by reviewing
how it can be used to describe a massive point particle.

\subsection{World-Line Description of a Point Particle}

A point particle sweeps out a trajectory (or world line) in spacetime.  This
can be described by functions $x^\mu (\tau)$ that describe how the world line,
parameterized by $\tau$, is embedded in the spacetime, whose coordinates are
denoted $x^\mu$.  For simplicity, let us assume that the spacetime is flat
Minkowski space with a Lorentz metric
\begin{equation}
\eta_{\mu\nu} = \left(\begin{array}{cccc}
-1 & 0 & 0 & 0\\
0 & 1 & 0 & 0\\
0 & 0 & 1 & 0\\
0 & 0 & 0 & 1 \end{array} \right).
\end{equation}
Then, the Lorentz invariant line element is given by
\begin{equation}
ds^2 = - \eta_{\mu\nu} dx^\mu dx^\nu.
\end{equation}
In units $\hbar = c = 1$, the action for a particle of mass $m$ is given by
\begin{equation}
S = - m \int ds.
\end{equation}
This could be generalized to a curved spacetime by replacing $\eta_{\mu\nu}$ by
a metric $g_{\mu\nu} (x)$, but we will not do so here.  In terms of the
embedding functions, $x^\mu (t)$, the action can be rewritten in the form
\begin{equation}
S = - m \int d\tau \sqrt{-\eta_{\mu\nu} \dot x^\mu \dot x^\nu},
\end{equation}
where dots represent $\tau$ derivatives.  An important property of this action
is invariance under local reparametrizations.
This is a kind of gauge invariance, whose
meaning is that the form of $S$ is unchanged under an arbitrary
reparametrization of the world line $\tau \rightarrow \tau (\tilde{\tau})$.
Actually, one should require that the function $\tau (\tilde{\tau})$ is smooth
and monotonic $\left(\frac{d\tau}{d\tilde{\tau}} > 0\right)$.  The
reparametrization invariance is a one-dimensional analog of the
four-dimensional general coordinate invariance of general relativity.
Mathematicians refer to this kind of symmetry as diffeomorphism invariance.

The reparametrization invariance of $S$ allows us to choose a gauge.  A nice
choice is the ``static gauge''
\begin{equation}
x^0 = \tau.
\end{equation}
In this gauge (renaming the parameter $t$) the action becomes
\begin{equation}
S = - m \int \sqrt{1 - v^2} dt,
\end{equation}
where
\begin{equation}
\vec v = \frac{d\vec x}{dt}.
\end{equation}
Requiring this action to be stationary under an arbitrary variation of $\vec x
(t)$ gives the Euler--Lagrange equations
\begin{equation}
\frac{d\vec p}{dt} = 0,
\end{equation}
where
\begin{equation}
\vec p = \frac{\delta S}{\delta \vec v} = \frac{m\vec v}{\sqrt{1 - v^2}},
\end{equation}
which is the usual result.  So we see that usual relativistic kinematics
follows from the action $S = - m \int ds$.

\subsection{World-Volume Actions}

We can now generalize the analysis of the massive point particle to a $p$-brane
of tension $T_p$.  The action in this case involves the invariant $(p +
1)$-dimensional volume and is given by
\begin{equation}
S_p = - T_p \int d\mu_{p + 1},
\end{equation}
where the invariant volume element is
\begin{equation}
d\mu_{p+1} = \sqrt{- \det (- \eta_{\mu\nu} \partial_\alpha x^\mu \partial_\beta
x^\nu)} d^{p+1} \sigma.
\end{equation}
Here the embedding of the $p$-brane into $d$-dimensional spacetime is given by
functions $x^\mu (\sigma^\alpha)$.  The index $\alpha = 0, \ldots, p$ labels
the $p + 1$ coordinates $\sigma^\alpha$ of the $p$-brane world-volume and the
index $\mu = 0, \ldots, d - 1$ labels the $d$ coordinates $x^\mu$ of the
$d$-dimensional spacetime.  We have defined
\begin{equation}
\partial_\alpha x^\mu = \frac{\partial x^\mu}{\partial\sigma^\alpha}.
\end{equation}
The determinant operation acts on the $(p+1)\times(p+1)$ matrix
whose rows and columns are labeled by $\alpha$ and $\beta$. The
tension $T_p$ is interpreted as the mass per unit volume of the
$p$-brane. For a 0-brane, it is just the mass.

\noindent{\bf Exercise:}  Show that $S_p$ is
reparametrization invariant.  In other words, substituting $\sigma^\alpha =
\sigma^\alpha (\tilde{\sigma}^\beta)$, it takes the same form when expressed in
terms of the coordinates $\tilde{\sigma}^\alpha$.

Let us now specialize to the string, $p = 1$.  Evaluating the determinant gives
\begin{equation}
S[x] = - T \int d\sigma d\tau \sqrt{\dot x^2 x^{\prime 2} - (\dot x \cdot
x')^2},
\end{equation}
where we have defined $\sigma^0 = \tau$, $\sigma^1 = \sigma$, and
\begin{equation}
\dot x^\mu = \frac{\partial x^\mu}{\partial\tau}, \quad x^{\prime\mu} =
\frac{\partial x^\mu}{\partial\sigma}.
\end{equation}
This action, called the Nambu--Goto action, was first proposed in
1970 \cite{Nambu,Goto}.  The Nambu--Goto action is equivalent to
the action
\begin{equation} \label{haction}
S[x,h] = - \frac{T}{2} \int d^2 \sigma \sqrt{-h} h^{\alpha\beta} \eta_{\mu\nu}
\partial_\alpha x^\mu \partial_\beta x^\nu,
\end{equation}
where $h_{\alpha\beta} (\sigma,\tau)$ is the world-sheet metric, $h = \det
h_{\alpha\beta}$, and $h^{\alpha\beta}$ is the inverse of $h_{\alpha\beta}$.
The Euler--Lagrange equation obtained by varying $h^{\alpha\beta}$ are
\begin{equation}
T_{\alpha\beta} = \partial_\alpha x \cdot \partial_\beta x - \frac{1}{2}
h_{\alpha\beta} h^{\gamma\delta} \partial_\gamma x \cdot \partial_\delta x = 0.
\end{equation}

\noindent {\bf Exercise:}  Show that $T_{\alpha\beta} = 0$ can be used to
eliminate the world-sheet metric from the action, and that when this is done
one recovers the Nambu--Goto action.  (Hint: take the determinant of both sides
of the equation $\partial_\alpha x \cdot \partial_\beta x = \frac{1}{2}
h_{\alpha\beta} h^{\gamma\delta} \partial_\gamma x \cdot \partial_\delta x.$)

In addition to reparametrization invariance, the action $S[x,h]$ has another
local symmetry, called conformal invariance (or Weyl invariance).
Specifically, it is invariant under the replacement
\begin{eqnarray}
h_{\alpha\beta} &\rightarrow& \Lambda (\sigma,\tau) h_{\alpha\beta}\\ \nonumber
x^\mu &\rightarrow& x^\mu.
\end{eqnarray}
This local symmetry is special to the $p = 1$ case (strings).

The two reparametrization invariance symmetries of $S[x,h]$ allow us to choose
a gauge in which the three functions $h_{\alpha\beta}$ (this is a symmetric
$2\times 2$ matrix) are expressed in terms of just one function.  A convenient
choice is the ``conformally flat gauge''
\begin{equation}
h_{\alpha\beta} = \eta_{\alpha\beta} e^{\phi(\sigma,\tau)}.
\end{equation}
Here, $\eta_{\alpha\beta}$ denoted the two-dimensional Minkowski metric of a
flat world sheet.  However, because of the factor $e^\phi, h_{\alpha\beta}$ is
only ``conformally flat.''  Classically, substitution of this gauge choice into
$S[x,h]$ leaves the gauge-fixed action
\begin{equation} \label{fixed}
S = \frac{T}{2} \int d^2 \sigma \eta^{\alpha\beta} \partial_\alpha x \cdot
\partial_\beta x.
\end{equation}

Quantum mechanically, the story is more subtle.  Instead of eliminating $h$ via
its classical field equations, one should perform a Feynman path integral,
using standard machinery to deal with the local symmetries and gauge
fixing.  When this is done correctly, one finds that in general $\phi$ does not
decouple from the answer.  Only for the special case $d = 26$ does the quantum
analysis reproduce the formula we have given based
on classical reasoning \cite{Polyakov}.
Otherwise, there are correction terms whose presence can be traced to a
conformal anomaly (i.e., a quantum-mechanical breakdown of the conformal
invariance).

The gauge-fixed action is quadratic in the $x$'s.  Mathematically,
it is the same as a theory of $d$ free scalar fields in two
dimensions. The equations of motion obtained by varying $x^\mu$
are simply free two-dimensional wave equations:
\begin{equation}
\ddot x^\mu - x^{\prime\prime\mu} = 0.
\end{equation}
This is not the whole story, however, because we must also take account of the
constraints $T_{\alpha\beta} = 0$.  Evaluated in the conformally flat gauge,
these constraints are
\begin{eqnarray}
T_{01} &=& T_{10} = \dot x \cdot x' = 0 \\ \nonumber
T_{00} &=& T_{11} = \frac{1}{2} (\dot x^2 + x^{\prime 2}) = 0.
\end{eqnarray}
Adding and subtracting gives
\begin{equation} \label{constraints}
(\dot x \pm x')^2 = 0.
\end{equation}

\subsection{Boundary Conditions}

To go further, one needs to choose boundary conditions.  There are three
important types.  For a closed string one should impose periodicity in the
spatial parameter $\sigma$.  Choosing its range to be $\pi$ (as is
conventional)
\begin{equation}
x^\mu (\sigma,\tau) = x^\mu (\sigma + \pi,\tau).
\end{equation}
For an open string (which has two ends), each end can be required to satisfy
either Neumann or Dirichlet boundary conditions (for each value of $\mu$).
\begin{eqnarray}
{\rm Neumann}: \quad \frac{\partial x^\mu}{\partial\sigma} &=& 0 \quad {\rm at} \,
\,
\sigma = 0 \,  \,{\rm or} \, \, \pi \\
{\rm Dirichlet}: \quad \frac{\partial x^\mu}{\partial\tau} &=& 0 \quad {\rm at} \,
\,
\sigma = 0 \, \,{\rm or} \, \, \pi.
\end{eqnarray}
The Dirichlet condition can be integrated, and then it specifies a
spacetime location on which the string ends.  The only way this
makes sense is if the open string ends on a physical object -- it
ends on a D-brane. (D stands for Dirichlet.) If all the
open-string boundary conditions are Neumann, then the ends of the
string can be anywhere in the spacetime.  The modern
interpretation is that this means that there are spacetime-filling
D-branes present.

Let us now consider the closed-string case in more detail.  The general
solution of the 2d wave equation is given by a sum of  ``right-movers'' and
``left-movers'':
\begin{equation}
x^\mu (\sigma, \tau) = x_R^\mu (\tau-\sigma) + x_L^\mu (\tau + \sigma).
\end{equation}
These should be subject to the following additional conditions:
\begin{itemize}
\item $x^\mu (\sigma,\tau)$ is real
\item $x^\mu (\sigma + \pi, \tau) = x^\mu (\sigma, \tau)$
\item $(x'_L)^2 = (x'_R)^2 = 0$ (These are the $T_{\alpha\beta} = 0$
constraints in eq. (\ref{constraints}).)
\end{itemize}
The first two of these conditions can be solved explicitly in terms of Fourier
series:
\begin{eqnarray}
x_R^\mu &=& \frac{1}{2} x^\mu + \ell_s^2 p^\mu (\tau - \sigma) +
\frac{i}{\sqrt{2}} \ell_s \sum_{n\not= 0} \frac{1}{n} \alpha_n^\mu
e^{-2in(\tau-\sigma)}\\ \nonumber
x_L^\mu &=& \frac{1}{2} x^\mu + \ell_s^2 p^\mu (\tau + \sigma) +
\frac{i}{\sqrt{2}} \ell_s \sum_{n\not= 0} \frac{1}{n} \tilde{\alpha}_n^\mu
e^{-2in(\tau+\sigma)},
\end{eqnarray}
where the expansion parameters $\alpha_n^\mu$,  $\tilde{\alpha}_n^\mu$ satisfy
\begin{equation}
\alpha_{-n}^\mu = (\alpha_n^{\mu})^{\dagger}, \quad
\tilde{\alpha}_{-n}^\mu = (\tilde{\alpha}_n^{\mu})^{\dagger}.
\end{equation}
The center-of-mass coordinate $x^\mu$ and momentum $p^\mu$ are also real.  The
fundamental string length scale $\ell_s$ is related to the tension $T$ by
\begin{equation}
T = \frac{1}{2\pi \alpha'}, \quad \alpha' = \ell_s^2.
\end{equation}
The parameter $\alpha'$ is called the universal Regge slope, since
the string modes lie on linear parallel Regge trajectories with
this slope.

\subsection{Quantization}

The analysis of closed-string left-moving modes, closed-string
right-moving modes, and open-string modes are all very similar.
Therefore, to avoid repetition, we will focus on the closed-string
right-movers.  Starting with the gauge-fixed action in
eq.(\ref{fixed}), the canonical momentum of the string is
\begin{equation}
p^\mu (\sigma, \tau) = \frac{\delta S}{\delta \dot x^\mu} = T \dot x^\mu.
\end{equation}
Canonical quantization (this is just free 2d field theory for scalar fields)
gives
\begin{equation}
[p^\mu (\sigma,\tau), x^\nu (\sigma',\tau)] = - i \hbar \eta^{\mu\nu} \delta
(\sigma - \sigma').
\end{equation}
In terms of the Fourier modes (setting $\hbar = 1$)
these become
\begin{equation}
[p^\mu, x^\nu] = - i\eta^{\mu\nu}
\end{equation}
\begin{eqnarray}
[\alpha_m^\mu, \alpha_n^\nu] &=& m \delta_{m + n, 0}
\eta^{\mu\nu},\\ \nonumber [\tilde\alpha_m^\mu,
\tilde\alpha_n^\nu] &=& m \delta_{m + n, 0} \eta^{\mu\nu},
\end{eqnarray}
and all other commutators vanish.

Recall that a quantum-mechanical harmonic oscillator can be described in terms
of raising and lowering operators, usually called $a^{\dagger}$ and $a$,
which satisfy
\begin{equation}
[a, a^{\dagger}] = 1.
\end{equation}
We see that, aside from a normalization factor, the expansion coefficients
$\alpha_{-m}^\mu$ and $\alpha_m^\mu$ are raising and lowering operators.  There
is just one problem.  Because $\eta^{00} = - 1$, the time components are
proportional to oscillators with the wrong sign $([a, a^{\dagger}] = - 1)$.  This is
potentially very bad, because such oscillators create states of negative norm,
which could lead to an inconsistent quantum theory (with negative
probabilities, etc.).  Fortunately, as we will explain, the $T_{\alpha\beta} =
0$ constraints eliminate the negative-norm states from the physical spectrum.

The classical constraint for the right-moving closed-string modes,
$(x'_R)^2 = 0$, has Fourier components
\begin{equation}
L_m = \frac{T}{2} \int_0^\pi e^{-2im\sigma} (x'_R)^2 d\sigma
= \frac{1}{2} \sum_{n=-\infty}^\infty \alpha_{m-n} \cdot \alpha_n,
\end{equation}
which are called Virasoro operators.  Since $\alpha_m^\mu$ does not commute
with $\alpha_{-m}^\mu$, $L_0$ needs to be normal-ordered:
\begin{equation}
L_0 = \frac{1}{2} \alpha_0^2 + \sum_{n=1}^\infty \alpha_{-n} \cdot \alpha_n.
\end{equation}
Here $\alpha_0^\mu = {\ell_s}p^\mu/{\sqrt{2}} $, where $p^\mu$ is the
momentum.

\subsection{The Free String Spectrum}

Recall that the Hilbert space of a harmonic oscillator is spanned by states
$|n\rangle, n = 0, 1,2, \ldots,$ where the ground state, $|0\rangle$, is
annihilated by the lowering operator $(a | 0\rangle = 0)$ and
\begin{equation}
|n\rangle = \frac{(a^{\dagger})^n}{\sqrt{n!}} |0\rangle.
\end{equation}
Then, for a normalized ground-state $(\langle 0|0\rangle = 1)$, one can use
$[a, a^{\dagger}] =1$ repeatedly to prove that
\begin{equation}
\langle m | n \rangle = \delta_{m,n}
\end{equation}
and
\begin{equation}
a^{\dagger} a |n\rangle = n | n \rangle.
\end{equation}
The string spectrum (of right-movers) is given by the product of an infinite
number of harmonic-oscillator Fock spaces, one for each $\alpha_n^\mu$, subject
to the Virasoro constraints \cite{Virasoro}
\begin{eqnarray}
(L_0 - q) |\phi \rangle &=& 0 \\ \nonumber
L_n |\phi \rangle &=& 0, \quad n > 0.
\end{eqnarray}
Here $|\phi\rangle$ denotes a physical state, and $q$ is a
constant to be determined.  It accounts for the arbitrariness in
the normal-ordering prescription used to define $L_0$.  As we will
see, the $L_0$ equation is a generalization of the Klein--Gordon
equation. It contains $p^2 = - \partial\cdot\partial$ plus
oscillator terms whose eigenvalue will determine the mass of the
state.

It is interesting to work out the algebra of the Virasoro operators $L_m$,
which follows from the oscillator algebra.  The result, called the Virasoro
algebra, is
\begin{equation}
[L_m, L_n] = (m-n) L_{m+n} + \frac{c}{12} (m^3 -m) \delta_{m+n,0}.
\end{equation}
The second term on the right-hand side is called the ``conformal anomaly term''
and the constant $c$ is called the ``central charge.''

\noindent{\bf Exercise:}  Verify
the first term on the right-hand side.  For extra credit, verify the second
term, showing that each component of $x^\mu$ contributes $c = 1$, so that
altogether $c = d$.

There are more sophisticated ways to describe the string spectrum
(in terms of BRST cohomology), but they are equivalent to the more
elementary approach presented here. In the BRST approach,
gauge-fixing to the conformal gauge in the quantum theory requires
the addition of world-sheet Faddeev-Popov ghosts, which turn out
to contribute $c = - 26$. Thus the total anomaly of the $x^\mu$
and the ghosts cancels for the particular choice $d = 26$, as we
asserted earlier.  Moreover, it is also necessary to set the
parameter $q = 1$, so that mass-shell condition becomes
\begin{equation}
(L_0 - 1) |\phi \rangle = 0.
\end{equation}

Since the mathematics of the open-string spectrum is the same as
that of closed-string right movers, let us now use the equations
we have obtained to study the open string spectrum.  (Here we are
assuming that the open-string boundary conditions are all Neumann,
corresponding to spacetime-filling D-branes.)  The mass-shell
condition is
\begin{equation}
M^2 = - p^2 = - \frac{1}{2} \alpha_0^2 = N - 1,
\end{equation}
where
\begin{equation}
N = \sum_{n = 1}^\infty \alpha_{-n} \cdot \alpha_n = \sum_{n=1}^\infty n
a^\dagger_n \cdot a_n.
\end{equation}
The $a^{\dagger}$'s and $a$'s are properly normalized raising and
lowering operators. Since each $a^{\dagger} a$ has eigenvalues
$0,1,2, \ldots$, the possible values of $N$ are also $0,1,2,
\ldots$.  The unique way to realize $N = 0$ is for all the
oscillators to be in the ground state, which we denote simply by
$|0;p^\mu\rangle$, where $p^\mu$ is the momentum of the state.
This state has $M^2 = - 1$, which is a tachyon ($p^\mu$ is
spacelike).  Such a faster-than-light particle is certainly not
possible in a consistent quantum theory, because the vacuum would
be unstable.  However, in perturbation theory (which is the
framework we are implicitly considering) this instability is not
visible. Since this string theory is only supposed to be a warm-up
exercise before considering tachyon-free superstring theories, let
us continue without worrying about it.

The first excited state, with $N=1$, corresponds to $M^2 = 0$.  The only way to
achieve $N = 1$ is to excite the first oscillator once:
\begin{equation}
|\phi \rangle = \zeta_\mu \alpha_{-1}^\mu |0;p\rangle.
\end{equation}
Here $\zeta_\mu$ denotes the polarization vector of a massless spin-one
particle.  The Virasoro constraint condition $L_1 |\phi\rangle = 0$ implies
that $\zeta_\mu$ must satisfy
\begin{equation}
p^\mu \zeta_\mu = 0.
\end{equation}
This ensures that the spin is transversely polarized, so there are $d-2$
independent polarization states.  This agrees with what one finds for a
massless Maxwell or Yang--Mills field.

At the next mass level, where $N=2$ and $M^2 = 1$, the most general possibility
has the form
\begin{equation}
|\phi\rangle = (\zeta_\mu \alpha_{-2}^\mu + \lambda_{\mu\nu} \alpha_{-1}^\mu
\alpha_{-1}^\nu)|0;p\rangle.
\end{equation}
However, the constraints $L_1 |\phi\rangle = L_2 |\phi\rangle = 0$
restrict $\zeta_\mu$ and $\lambda_{\mu\nu}$.  The analysis is
interesting, but only the results will be described.  If $d > 26$,
the physical spectrum contains a negative-norm state, which is not
allowed. However, when $d=26$, this state becomes zero norm and
decouples from the theory.  This leaves a pure massive ``spin
two'' (symmetric traceless tensor) particle as the only physical
state at this mass level.

Let us now turn to the closed-string spectrum.  A closed-string state is
described as a tensor product of a left-moving state and a right-moving state,
subject to the condition that the $N$ value of the left-moving and the right-moving
state is the same.  The reason for this ``level-matching'' condition is that we
have $(L_0 - 1) |\phi\rangle = (\tilde{L}_0 - 1) |\phi\rangle = 0$.  The sum
$(L_0 + \tilde{L}_0 - 2)|\phi\rangle$ is interpreted as the mass-shell
condition, while the difference $(L_0 - \tilde{L}_0)|\phi\rangle = (N -
\tilde{N}) |\phi\rangle = 0$ is the level-matching condition.

Using this rule, the closed-string ground state is just
\begin{equation}
|0\rangle \otimes |0\rangle,
\end{equation}
which represents a spin $0$ tachyon with $M^2 = - 2$. (The
notation no longer displays the momentum $p$ of the state.) Again,
this signals an unstable vacuum, but we will not worry about it
here. Much more important, and more significant, is the first
excited state
\begin{equation}
|\phi\rangle = \zeta_{\mu\nu} (\alpha_{-1}^\mu |0\rangle \otimes
\tilde{\alpha}_{-1}^\nu |0\rangle),
\end{equation}
which has $M^2 = 0$.  The Virasoro constraints $L_1 |\phi\rangle =
\tilde{L}_1 |\phi\rangle = 0$ imply that $p^\mu \zeta_{\mu\nu} =
0$.  Such a polarization tensor encodes three distinct spin
states, each of which plays a fundamental role in string theory.
The symmetric part of $\zeta_{\mu\nu}$ encodes a spacetime metric
field $g_{\mu\nu}$ (massless spin two) and a scalar dilaton field
$\phi$ (massless spin zero).  The $g_{\mu\nu}$ field is the
graviton field, and its presence (with the correct gauge
invariances) accounts for the fact that the theory contains
general relativity, which is a good approximation for $E\ll
1/\ell_s$. Its vacuum value determines the spacetime geometry.
Similarly, the value of $\phi$ determines the string coupling
constant ($g_s =<e^{\phi}>$).

$\zeta_{\mu\nu}$ also has an antisymmetric part, which corresponds to a
massless antisymmetric tensor gauge field $B_{\mu\nu} = - B_{\nu\mu}$.  This
field has a gauge transformation of the form
\begin{equation}
\delta B_{\mu\nu} = \partial_\mu \Lambda_\nu - \partial_\nu \Lambda_\mu,
\end{equation}
(which can be regarded as a generalization of the gauge
transformation rule for the Maxwell field: $\delta A_\mu =
\partial_\mu \Lambda)$. The gauge-invariant field strength
(analogous to $F_{\mu\nu} =
\partial_\mu A_\nu - \partial_\nu A_\mu)$ is
\begin{equation}
H_{\mu\nu\rho} = \partial_\mu B_{\nu\rho} + \partial_\nu B_{\rho\mu} +
\partial_\rho B_{\mu\nu}.
\end{equation}
The importance of the $B_{\mu\nu}$ field resides in the fact that the
fundamental string is a source for $B_{\mu\nu}$, just as a charged particle is
a source for the vector potential $A_\mu$.  Mathematically, this is expressed
by the coupling
\begin{equation}
q \int B_{\mu\nu} dx^\mu \wedge dx^\nu,
\end{equation}
which generalizes the coupling of a charged particle to a Maxwell
field
\begin{equation}
q \int A_\mu dx^\mu
\end{equation}
in a convenient notation.

\subsection{The Number of Physical States}

The number of physical states grows rapidly as a function of mass.
This can be analyzed quantitatively.  For the open string, let us
denote the number of physical states with $\alpha' M^2 = n - 1$ by
$d_n$. These numbers are encoded in the generating function
\begin{equation}
G(w) = \sum_{n = 0}^\infty d_n w^n  = \prod_{m = 1}^\infty (1 - w^m)^{-24}.
\end{equation}
The exponent 24 reflects the fact that in 26 dimensions, once the Virasoro
conditions are taken into account, the spectrum is exactly what one would get
from 24 transversely polarized oscillators.  It is easy to deduce from this
generating function the asymptotic number of states for large $n$, as a
function of $n$
\begin{equation}
d_n \sim n^{-27/4} e^{4\pi\sqrt{n}}.
\end{equation}

\noindent{\bf Exercise:}  Verify this formula.

\noindent This asymptotic degeneracy implies that the
finite-temperature partition function
\begin{equation}
{\rm tr}\, (e^{-\beta H}) = \sum_{n=0}^{\infty} d_n e^{-\beta M_n}
\end{equation}
diverges for $\beta^{-1} = T > T_H$, where $T_H$ is the Hagedorn
temperature
\begin{equation}
T_H = \frac{1}{4\pi\sqrt{\alpha'}} = \frac{1}{4\pi\ell_s} .
\end{equation}
$T_H$ might be the maximum possible temperature or else a critical temperature
at which there is a phase transition.

\subsection{The Structure of String Perturbation Theory}

As we discussed in the first lecture, perturbation theory
calculations are carried out by computing Feynman diagrams.
Whereas in ordinary quantum field theory Feynman diagrams are webs
of world lines, in the case of string theory they are
two-dimensional surfaces representing string world sheets. For
these purposes, it is convenient to require that the world-sheet
geometry is Euclidean (i.e., the world-sheet metric
$h_{\alpha\beta}$ is positive definite). The diagrams are
classified by their topology, which is very well understood in the
case of two-dimensional surfaces. The world-sheet topology is
characterized by the number of handles $(h)$, the number of
boundaries $(b)$, and whether or not they are orientable.  The
order of the expansion (i.e., the power of the string coupling
constant) is determined by the Euler number of the world sheet
$M$.  It is given by $\chi(M) = 2 - 2h - b$.  For example, a
sphere has $h = b = 0$, and hence $\chi = 2$. A torus has $h = 1$,
$b = 0$, and $\chi = 0$, a cylinder has $h = 0$, $b = 2$, and
$\chi = 0$, and so forth. Surfaces with $\chi = 0$ admit a flat
metric.

A scattering amplitude is given by a path integral of the schematic structure
\begin{equation}
\int Dh_{\alpha\beta} (\sigma) Dx^\mu (\sigma) e^{-S[h,x]}\prod_{i =
1}^{n_c} \int_M V_{\alpha_{i}} (\sigma_i) d^2 \sigma_i
\prod_{j=1}^{n_o}\int_{\partial M} V_{\beta_j} (\sigma_j) d\sigma_j.
\end{equation}
The action $S[h,x]$ is given in eq.~(\ref{haction}).
$V_{\alpha_{i}}$ is a vertex operator that describes emission or
absorption of a closed-string state of type $\alpha_i$ from the
interior of the string world sheet, and $V_{\beta_{j}}$ is a
vertex operator that describes emission of absorption of an
open-string state of type $\beta_j$ from the boundary of the
string world sheet.  There are lots of technical details that are
not explained here. In the end, one finds that the conformally
inequivalent world sheets of a given topology are described by a
finite number of parameters, and thus these amplitudes can be
recast as finite-dimensional integrals over these ``moduli.'' (The
momentum integrals are already done.) The dimension of the
resulting integral turns out to be
\begin{equation}
N = 3 (2h + b - 2) + 2n_c + n_o.
\end{equation}

As an example consider the amplitude describing elastic scattering of two-open
string ground states.  In this case $h = 0$, $b = 1$, $n_c = 0$, $n_o = 4$, and
therefore $N = 1$.  In terms of the usual Mandelstam invariants $s = - (p_1 +
p_2)^2$ and $t = - (p_1 - p_4)^2$, the result is
\begin{equation}
A(s,t) = g_s^2 \int_0^1 dx~~x^{-\alpha(s)-1} (1 - x)^{-\alpha(t)-1},
\end{equation}
where the Regge trajectory $\alpha(s)$ is
\begin{equation}
\alpha(s) = 1 + \alpha' s.
\end{equation}
This integral is just the Euler beta function
\begin{equation}
A(s,t) = g_s^2 B (-\alpha (s), - \alpha (t))
= g_s^2 \frac{\Gamma (-\alpha (s))
\Gamma (-\alpha (t))}{\Gamma (-\alpha (s) - \alpha (t))}.
\end{equation}
This is the famous Veneziano amplitude \cite{Veneziano}, which got the whole
business started.

\subsection{Recapitulation}

This lecture described some of the basic facts of the
26-dimensional bosonic string theory. One significant point that
has not yet been made clear is that there are actually a number of
distinct theories depending on what kinds of strings one includes

\begin{itemize}
\item oriented closed strings only
\item oriented closed strings and oriented open strings.  In this case one can
incorporate $U(n)$ gauge symmetry.
\item unoriented closed strings only
\item unoriented closed strings and unoriented open strings.  In this case one
can incorporate $SO(n)$ or $Sp(n)$ gauge symmetry.
\end{itemize}

As we have mentioned already, all the bosonic string theories are
sick as they stand, because (in each case) the closed-string
spectrum contains a tachyon.  A tachyon means that one is doing
perturbation theory about an unstable vacuum.  This is analogous
to the unbroken symmetry extremum of the Higgs potential in the
standard model. In that case, we know that there is a stable
minimum, where the Higgs fields acquires a vacuum value. It is
conceivable that the closed-string tachyon condenses in an
analogous manner, or else there might not be a stable vacuum.
Recently, there has been success in demonstrating that open-string
tachyons condense at a stable minimum, but the fate of
closed-string tachyons is still an open problem.

\section{Lecture 3: Superstrings}

Among the deficiencies of the bosonic string theory is the fact
that there are no fermions.  As we will see, the addition of
fermions leads quite naturally to supersymmetry and hence
superstrings.  There are two alternative formalisms that are used
to study superstrings.  The original one, which grew out of the
1971 papers by Ramond \cite{Ramond} and by Neveu and me
\cite{Neveu}, is called the RNS formalism.  In this approach, the
supersymmetry of the two-dimensional world-sheet theory plays a
central role.  The second approach, developed by Michael Green and
me in the early 1980's \cite{Greena}, emphasizes supersymmetry in
the ten-dimensional spacetime.  Due to lack of time, only the RNS
approach will be presented.

In the RNS formalism, the world-sheet theory is based on the $d$
functions $x^\mu (\sigma, \tau)$ that describe the embedding of
the world sheet in the spacetime, just as before.  However, in
order to supersymmetrize the world-sheet theory, we also introduce
$d$ fermionic partner fields $\psi^\mu (\sigma,\tau)$.  Note that
$x^\mu$ transforms as a vector from the spacetime viewpoint, but
as $d$ scalar fields from the two-dimensional world-sheet
viewpoint. The $\psi^\mu$ also transform as a spacetime vector,
but as world-sheet spinors. Altogether, $x^\mu$ and $\psi^\mu$
described $d$ supersymmetry multiplets, one for each value of
$\mu$.

The reparametrization invariant world-sheet action described in
the preceding lecture can be generalized to have local
supersymmetry on the world sheet, as well.  (The details of how
that works are a bit too involved to describe here.) When one
chooses a suitable conformal gauge $(h_{\alpha\beta} = e^\phi
\eta_{\alpha\beta})$, together with an appropriate fermionic gauge
condition, one ends up with a world-sheet theory that has global
supersymmetry supplemented by constraints.  The constraints form a
super-Virasoro algebra. This means that in addition to the
Virasoro constraints of the bosonic string theory, there are
fermionic constraints, as well.

\subsection{The Gauge-Fixed Theory}

The globally supersymmetric world-sheet action that arises in the
conformal gauge takes the form
\begin{equation} \label{susyfixed}
S = - \frac{T}{2} \int d^2 \sigma (\partial_\alpha x^\mu
\partial^\alpha x_\mu - i \bar \psi^\mu \rho^\alpha
\partial_\alpha \psi_\mu).
\end{equation}
The first term is exactly the same as in eq.~(\ref{fixed}) of the
bosonic string theory. Recall that it has the structure of $d$
free scalar fields.  The second term that has now been added is
just $d$ free massless spinor fields, with Dirac-type actions. The
notation is that $\rho^\alpha$ are two $2\times 2$ Dirac matrices
and $\psi = \binom{\psi_-}{\psi_+}$ is a two-component Majorana
spinor.  The Majorana condition simply means that $\psi_+$ and
$\psi_-$ are real in a suitable representation of Dirac algebra.
In fact, a convenient choice is one for which
\begin{equation}
\bar \psi \rho^\alpha \partial_\alpha \psi =
\psi_- \partial_+ \psi_- + \psi_+
\partial_- \psi_+,
\end{equation}
where $\partial_\pm$ represent derivatives with respect to $\sigma^\pm = \tau
\pm \sigma$.  In this basis, the equations of motion are simply
\begin{equation}
\partial_+ \psi^\mu_- = \partial_- \psi_+^\mu = 0.
\end{equation}
Thus $\psi_-^\mu$ describes right-movers and $\psi_+^\mu$ describes
left-movers.

Concentrating on the right-movers $\psi_-^\mu$, the global supersymmetry
transformations, which are a symmetry of the gauge-fixed action, are
\begin{eqnarray}
\delta x^\mu &=& i \epsilon \psi_-^\mu\\ \nonumber
\delta \psi_-^\mu &=& - 2 \partial_- x^\mu \epsilon.
\end{eqnarray}

\noindent {\bf Exercise:} Show that this is a symmetry of the
action (\ref{susyfixed}).

\noindent There is an analogous symmetry for the left-movers.
(Accordingly, the world-sheet theory is said to have $(1,1)$
supersymmetry.) Continuing to focus on the right-movers, the
Virasoro constraint is
\begin{equation}
 (\partial_- x)^2 + \frac{i}{2} \psi_-^\mu \partial_- \psi_{\mu-}
= 0.
\end{equation}
The first term is what we found in the bosonic string theory, and
the second term is an additional fermionic contribution.  There is
also an associated fermionic constraint
\begin{equation}
\psi_-^\mu \partial_- x_\mu = 0.
\end{equation}
The Fourier modes of these constraints satisfy the super-Virasoro
algebra. There is a second identical super-Virasoro algebra for
the left-movers.

As in the bosonic string theory, the Virasoro algebra has
conformal anomaly terms proportional to a central charge $c$.  As
in that theory, each component of $x^\mu$ contributes $+1$ to the
central charge, for a total of $d$, while (in the BRST
quantization approach) the reparametrization symmetry ghosts
contribute $-26$. But now there are additional contributions. Each
component of $\psi^\mu$ gives $+1/2$, for a total of $d/2$, and
the local supersymmetry ghosts contribute $+11$.  Adding all of
this up, gives a grand total of $c = \frac{3d}{2} - 15$.  Thus, we
see that the conformal anomaly cancels for the specific choice $d
= 10$.  This is the preferred critical dimension for superstrings,
just as $d = 26$ is the critical dimension for bosonic strings.
For other values the theory has a variety of inconsistencies.

\subsection{The R and NS Sectors}

Let us now consider boundary conditions for $\psi^\mu (\sigma, \tau)$.  (The
story for $x^\mu$ is exactly as before.)  First, let us consider open-string
boundary conditions.  For the action to be well-defined, it turns out that one
must set $\psi_+ = \pm \psi_-$ at the two ends $\sigma = 0, \pi$.  An overall
sign is a matter of convention, so we can set
\begin{equation}
\psi_+^\mu (0,\tau) = \psi_-^\mu (0,\tau),
\end{equation}
without loss of generality.  But this still leaves two possibilities for the
other end, which are called R and NS:
\begin{eqnarray}
{\rm R} &:& \psi_+^\mu (\pi,\tau) = \psi_-^\mu (\pi,\tau)\\ \nonumber
{\rm NS} &:& \psi_+^\mu (\pi,\tau) = -\psi_-^\mu (\pi,\tau).
\end{eqnarray}
Combining these with the equations of motion $\partial_- \psi_+ = \partial_+
\psi_- = 0$, allows us to express the general solutions as Fourier series
\begin{eqnarray}
{\rm R}:\quad \psi_-^\mu &=& \frac{1}{\sqrt{2}} \sum_{n\in {\bf
Z}} d_n^\mu e^{-in (\tau - \sigma)}\\ \nonumber \psi_+^\mu &=&
\frac{1}{\sqrt{2}} \sum_{n\in {\bf Z}} d_n^\mu e^{-in (\tau +
\sigma)}\\ \nonumber {\rm NS}:\quad \psi_-^\mu &=&
\frac{1}{\sqrt{2}} \sum_{r\in {\bf Z} + 1/2} b_r^\mu e^{-ir (\tau
- \sigma)}\\ \nonumber \psi_+^\mu &=& \frac{1}{\sqrt{2}}
\sum_{r\in {\bf Z} + 1/2} b_r^\mu e^{-ir (\tau + \sigma)}.
\end{eqnarray}
The Majorana condition implies that $d_{-n}^\mu =
d_n^{\mu\dagger}$ and $b_{-r}^\mu = b_r^{\mu\dagger}$.  Note that
the index $n$ takes integer values, whereas the index $r$ takes
half-integer values ($\pm \frac{1}{2}, \pm \frac{3}{2}, \ldots$).
In particular, only the R boundary condition gives a zero mode.

Canonical quantization of the free fermi fields $\psi^\mu (\sigma, \tau)$ is
very standard and straightforward.  The result can be expressed as
anticommutation relations for the coefficients $d_m^\mu$ and $b_r^\mu$:
\begin{eqnarray}
{\rm R} &:& \qquad \{d_n^\mu, d_n^\nu\} = \eta^{\mu\nu} \delta_{m+n,0} \qquad m,n
\in {\bf Z}\\ \nonumber
{\rm NS} &:& \qquad \{d_r^\mu, d_s^\nu\} = \eta^{\mu\nu} \delta_{r+s,0} \qquad r,s
\in {\bf Z} + \frac{1}{2}.
\end{eqnarray}
Thus, in addition to the harmonic oscillator operators
$\alpha_m^\mu$ that appear as coefficients in mode expansions of
$x^\mu$, there are fermionic oscillator operators  $d_m^\mu$ or
$b_r^\mu$ that appear as coefficients in mode expansions of
$\psi^\mu$.  The basic structure $\{b,b^\dagger\} = 1$ is very
simple.  It describes a two-state system with $b|0\rangle = 0$,
and $b^\dagger|0\rangle = |1\rangle$.  The $b$'s or $d$'s with
negative indices can be regarded as raising operators and those
with positive indices as lowering operators, just as we did for
the $\alpha_n^\mu$.

In the NS sector, the ground state $|0; p\rangle$ satisfies
\begin{equation}
\alpha_m^\mu |0;p\rangle = b_r^\mu |0;p\rangle = 0, \quad m,r > 0
\end{equation}
which is a straightforward generalization of how we defined the
ground state in the bosonic string theory.  All the excited states
obtained by acting with the $\alpha$ and $b$ raising operators are
spacetime bosons.  We will see later that the ground state,
defined as we have done here, is again a tachyon. However, in this
theory, as we will also see, there is a way by which this tachyon
can (and must) be removed from the physical spectrum.

In the R  sector there are zero modes that satisfy the algebra
\begin{equation}
\{d_0^\mu, d_0^\nu\} = \eta^{\mu\nu}.
\end{equation}
This is the $d$-dimensional spacetime Dirac algebra.  Thus the $d_0$'s should be
regarded as Dirac matrices and all states in the R sector should be spinors
in order to furnish representation spaces on which these operators can act.  The
conclusion, therefore, is that whereas all string states in the NS sector are
spacetime bosons, all string states in the R sector are spacetime fermions.

In the closed-string case, the physical states are obtained by
tensoring right-movers and left-movers, each of which are
mathematically very similar to the open-string spectrum.  This
means that there are four distinct sectors of closed-string
states:  NS$\otimes$NS and R$\otimes$R describe spacetime bosons,
whereas NS$\otimes$R and R$\otimes$NS describe spacetime fermions.
We will return to explore what this gives later, but first we need
to explore the right-movers by themselves in more detail.

The zero mode of the fermionic constraint $\psi^\mu \partial_-
x_\mu = 0$ gives a wave equation for (fermionic) strings in the
Ramond sector,  $F_0 |\psi\rangle = 0$, which is called the
Dirac--Ramond equation.  In terms of the oscillators
\begin{equation}
F_0 = \alpha_0 \cdot d_0 + \sum_{n\not= 0} \alpha_{-n} \cdot d_n.
\end{equation}
The zero-mode piece of $F_0$, $\alpha_0 \cdot d_0$, has been
isolated, because it is just the usual Dirac operator, $\gamma^\mu
\partial_\mu$, up to normalization. (Recall that $\alpha_{0 \mu}$
is proportional to $p_\mu = - i\partial_\mu$, and $d_0^\mu$ is
proportional to the Dirac matrices $\gamma^\mu$.)  The fermionic
ground state $|\psi_0\rangle$, which satisfies
\begin{equation}
\alpha_n^\mu |\psi_0\rangle = d_n^\mu |\psi_0\rangle = 0, \quad n > 0,
\end{equation}
satisfies the wave equation
\begin{equation}
\alpha_0 \cdot d_0 |\psi_0 \rangle = 0,
\end{equation}
which is precisely the massless Dirac equation.  Hence the fermionic ground
state is a massless spinor.

\subsection{The GSO Projection}

In the NS (bosonic) sector the mass formula is
\begin{equation}
M^2 = N - \frac{1}{2},
\end{equation}
which is to be compared with the formula $M^2 = N - 1$ of the
bosonic string theory.  This time the number operator $N$ has
contributions from the $b$ oscillators as well as the $\alpha$
oscillators. (The reason that the normal-ordering constant is
$-1/2$ instead of $-1$ works as follows. Each transverse $\alpha$
oscillator contributes $-1/24$ and each transverse $b$ oscillator
contributes $-1/48$. The result follows since the bosonic theory
has 24 transverse directions and the superstring theory has 8
transverse directions.) Thus the ground state, which has $N = 0$,
is now a tachyon with $M^2 = - 1/2$.

This is where things stood until the 1976 work of Gliozzi, Scherk,
and Olive \cite{GSO}.  They noted that the spectrum admits a
consistent truncation (called the GSO projection) which is
necessary for the consistency of the interacting theory.  In the
NS sector, the GSO projection keeps states with an odd number of
$b$-oscillator excitations, and removes states with an even number
of $b$-oscillator excitation.  Once this rule is implemented the
only possible values of $N$ are half integers, and the spectrum of
allowed masses are integral
\begin{equation}
M^2 = 0,1,2, \ldots.
\end{equation}
In particular, the bosonic ground state is now massless.  The
spectrum no longer contains a tachyon. The GSO projection also
acts on the R sector, where there is an analogous restriction on
the $d$ oscillators.  This amounts to imposing a chirality
projection on the spinors.

Let us look at the massless spectrum of the GSO-projected theory.
The ground state boson is now a massless vector, represented by
the state $\zeta_\mu b_{-1/2}^\mu |0;p\rangle$, which (as before)
has $d - 2 = 8$ physical polarizations.  The ground state fermion
is a massless Majorana--Weyl fermion which has $\frac{1}{4} \cdot
2^{d/2} = 8$ physical polarizations.  Thus there are an equal
number of bosons and fermions, as is required for a theory with
spacetime supersymmetry.  In fact, this is the pair of fields that
enter into ten-dimensional super Yang--Mills theory.  The claim is
that the complete theory now has spacetime supersymmetry.

If there is spacetime supersymmetry, then there should be an equal
number of bosons and fermions at every mass level.  Let us denote
the number of bosonic states with $M^2 = n$ by $d_{NS}(n)$ and the
number of fermionic states with $M^2 = n$ by $d_R(n)$.  Then we
can encode these numbers in generating functions, just as we did
for the bosonic string theory
\begin{equation}
f_{NS}(w) = \sum_{n=0}^\infty d_{NS} (n) w^n =\frac{1}{2\sqrt{w}}
\left(\prod_{m=1}^\infty \left(\frac{1+w^{m-1/2}}{1 -
w^m}\right)^8 - \prod_{m=1}^\infty \left(\frac{1 -
w^{m-1/2}}{1-w^m}\right)^8\right)
\end{equation}
\begin{equation}
 f_{R}(w) =
\sum_{n=0}^\infty d_R (n) w^n = 8 \prod_{m=1}^\infty
\left(\frac{1+w^m}{1-w^m}\right)^8.
\end{equation}
The $8$'s in the exponents refer to the number of transverse directions in ten
dimensions.  The effect of the GSO projection is the subtraction of the second
term in $f_{NS}$ and reduction of coefficient in $f_R$ from 16 to 8.  In 1829,
Jacobi discovered the formula
\begin{equation}
f_R (w) = f_{NS} (w).
\end{equation}
(He used a different notation, of course.) For him this relation
was an obscure curiosity, but we now see that it provides strong
evidence for supersymmetry of this string theory in ten
dimensions.  A complete proof of supersymmetry for the interacting
theory was constructed by Green and me five years after the GSO
paper~\cite{Greena}.

\subsection{Type II Superstrings}

We have described the spectrum of bosonic (NS) and fermionic (R)
string states. This also gives the spectrum of left-moving and
right-moving closed-string modes, so we can form the closed-string
spectrum by forming tensor products as before. In particular, the
massless right-moving spectrum consists of a vector and a
Majorana--Weyl spinor.  Thus the massless closed-string spectrum
is given by
\begin{equation}
({\rm vector}~ + {\rm MW} ~{\rm spinor}) ~ \otimes ~({\rm vector}
~ + {\rm MW} ~{\rm spinor}).
\end{equation}
There are actually two distinct possibilities because two MW
spinor can have either opposite chirality or the same chirality.

When the two MW spinors have opposite chirality, the theory is
called type IIA superstring theory, and its massless spectrum
forms the type IIA supergravity multiplet.  This theory is
left-right symmetric.  In other words, the spectrum is invariant
under mirror reflection.  This implies that the IIA theory is
parity conserving. When the two MW spinors have the same
chirality, the resulting type IIB superstring theory is chiral,
and hence parity violating. In each case there are two gravitinos,
arising from vector $\otimes$ spinor and spinor $\otimes$ vector,
which are gauge fields for local supersymmetry. (In four
dimensions we would say that the gravitinos have spin $3/2$, but
that is not an accurate description in ten dimensions.)  Thus,
since both type II superstring theories have two gravitinos, they
have local ${\cal N} = 2$ supersymmetry in the ten-dimensional
sense. The supersymmetry charges are Majorana--Weyl spinors, which
have 16 components, so the type II theories have 32 conserved
supercharges.  This is the same amount of supersymmetry as what is
usually called ${\cal N} = 8$ in four dimensions.

The type II superstring theories contain only oriented closed
strings (in the absence of D-branes).  However, there is another
superstring theory, called type I, which can be obtained by a
projection of the type IIB theory, that only keeps the diagonal
sum of the two gravitinos.  Thus, this theory only has ${\cal N} =
1$ supersymmetry (16 supercharges).  It is a theory of unoriented
closed strings.  However, it can be supplemented by unoriented
open strings. This introduces a Yang--Mills gauge group, which
classically can be $SO(n)$ or $Sp(n)$ for any value of $n$.
Quantum consistency singles out $SO(32)$ as the unique
possibility.  This restriction can be understood in a number of
ways. The way that it was first discovered was by considering
anomalies.

\subsection{Anomalies}

Chiral (parity-violating) gauge theories can be inconsistent due
to anomalies. This happens when there is a quantum mechanical
breakdown of the gauge symmetry, which is induced by certain
one-loop Feynman diagrams. (Sometimes one also considers breaking
of global symmetries by anomalies, which does not imply an
inconsistency. That is not what we are interested in here.) In the
case of four dimensions, the relevant diagrams are triangles, with
the chiral fields going around the loop and three gauge fields
attached as external lines. In the case of the standard model, the
quarks and leptons are chiral and contribute to a variety of
possible anomalies. Fortunately, the standard model has just the
right content so that all of the gauge anomalies cancel. If one
discarded the quark or lepton contributions, it would not work.

In the case of ten-dimensional chiral gauge theories, the potentially anomalous
Feynman diagrams are hexagons, with six external gauge fields.  The anomalies
can be attributed to the massless fields, and therefore they can be analyzed in
the low-energy effective field theory.  There are several possible cases in ten
dimensions:

\begin{itemize}
\item ${\cal N} = 1$ supersymmetric Yang--Mills theory.  This theory has
anomalies for every choice of gauge group.
\item Type I supergravity.  This theory has gravitational anomalies.
\item Type IIA supergravity.  This theory is non-chiral, and therefore it is
trivially anomaly-free.
\item Type IIB supergravity.  This theory has three chiral fields each of which
contributes to several kinds of gravitational anomalies.  However, when their
contributions are combined, the anomalies all cancel.  (This result was
obtained by Alvarez--Gaum\'e and Witten in 1983 \cite{Alvarez}.)
\item Type I supergravity coupled to super Yang--Mills.  This theory has both
gauge and gravitational anomalies for every choice of Yang-Mills gauge group
except $SO(32)$ and $E_8 \times E_8$.  For these two choices, all the anomalies
cancel.  (This result was obtained by Green and me in 1984 \cite{Greenb}.)
\end{itemize}

As we mentioned earlier, at the classical level one can define
type I superstring theory for any orthogonal or symplectic gauge
group.  Now we see that at the quantum level, the only choice that
is consistent is $SO(32)$.  For any other choice there are fatal
anomalies.  The term $SO(32)$ is used here somewhat imprecisely.
There are several different Lie groups that have the same Lie
algebra. It turns out that the precise Lie group that is
appropriate is Spin (32)$/{\bf Z}_2$.

\subsection{Heterotic Strings}

The two Lie groups that are singled out --- $E_8 \times E_8$
and Spin (32)$/{\bf Z}_2$ ---
have several properties in common.  Each of them has dimension $= 496$ and rank
$= 16$.  Moreover, their weight lattices correspond to the only two even
self-dual lattices in 16 dimensions.  This last fact was the crucial clue that
led Gross, Harvey, Martinec, and Rohm \cite{Gross} to the
discovery of the heterotic string
soon after the anomaly cancellation result.  One hint is the relation
$10+16=26$.  The construction of the heterotic string uses the $d=26$ bosonic
string for the left-movers and the $d=10$ superstring the right movers.  The
sixteen extra left-moving dimensions are associated to an even self-dual
16-dimensional lattice.  In this way one builds in the $SO(32)$ or $E_8 \times
E_8$ gauge symmetry.

Thus, to recapitulate, by 1985 we had five consistent superstring
theories, type I (with gauge group $SO(32)$), the two type II
theories, and the two heterotic theories.  Each is a
supersymmetric ten-dimensional theory.  The perturbation theory
was studied in considerable detail, and while some details may not
have been completed, it was clear that each of the five theories
has a well-defined, ultraviolet-finite perturbation expansion,
satisfying all the usual consistency requirements (unitarity,
analyticity, causality, etc.)  This was pleasing, though it was
somewhat mysterious why there should be five consistent quantum
gravity theories.  It took another ten years until we understood
that these are actually five special quantum vacua of a unique
underlying theory.

\subsection{T Duality}

T duality, an amazing result obtained in the late 1980's, relates one string
theory with a circular compact dimension of radius $R$ to another string theory
with a circular dimension of radius $1/R$ (in units $\ell_s = 1$).  This is
very profound, because it indicates a limitation of our usual motions of
classical geometry.  Strings see geometry differently from point particles.
Let us examine how this is possible.

The key to understanding T duality is to consider the kinds of
excitations that a string can have in the presence of a circular
dimension.  One class of excitations, called Kaluza--Klein
excitations, is a very general feature of any quantum theory,
whether or not based on strings.  The idea is that in order for
the wave function $e^{ipx}$ to be single valued, the momentum
along the circle must be a multiple of $1/R$, $p = n/R$, where $n$
is an integer.  From the lower-dimension viewpoint this is
interpreted as a contribution $(n/R)^2$ to the square of the mass.

There is a second type of excitation that is special to closed
strings. Namely, a closed string can wind $m$ times around the
circular dimension, getting caught up on the topology of the
space, contributing an energy given by the string tension times
the length of the string
\begin{equation}
E_m = 2\pi R \cdot m \cdot T.
\end{equation}
Putting $T = \frac{1}{2\pi}$ (for $\ell_s = 1$), this is just $E_m
= m R$.

The combined energy-squared of the Kaluza--Klein and winding-mode
excitations is
\begin{equation}
E^2 = \left(\frac{n}{R}\right)^2 + (m R)^2 + \ldots,
\end{equation}
where the dots represent string oscillator contributions.  Under T duality
\begin{equation}
m \leftrightarrow n, ~~~R \leftrightarrow 1/R.
\end{equation}
Together, these interchanges leave the energy invariant. This
means that what is interpreted as a Kaluza--Klein excitation in
one string theory is interpreted as a winding-mode excitation in
the T-dual theory, and the two theories have radii $R$ and $1/R$,
respectively.  The two principle examples of T-dual pairs are the
two type II theories and the two heterotic theories. In the latter
case there are additional technicalities that explain how the two
gauge groups are related. Basically, when the compactification on
a circle to nine dimension is carried out in each case, it is
necessary to include effects that we haven't explained (called
Wilson lines) to break the gauge groups to $SO(16) \times SO(16)$,
which is a common subgroup of $SO(32)$ and $E_8 \times E_8$.

\section{Lecture 4: From Superstrings to M Theory}

Superstring theory is currently undergoing a period of rapid development in which
important advances in understanding are being achieved.
The focus in this lecture will be on explaining why there can be an
eleven-dimensional vacuum, even though there are
only ten dimensions in perturbative
superstring theory.  The nonperturbative
extension of superstring theory that allows for an
eleventh dimension has been named {\em M theory}.  The letter M is intended to
be flexible in its interpretation.  It could stand for {\em magic,} {\em mystery,}
or {\em meta} to reflect our current state of incomplete understanding.  Those
who think that two-dimensional supermembranes (the M2-brane) are fundamental
may regard M as standing for {\em membrane.}  An approach called {\em Matrix
theory}  is another possibility.  And, of course, some view M theory as the
{\em mother} of all theories.

In the first superstring revolution we identified five distinct
superstring theories, each in ten dimensions.  Three of them, the
{type I} theory and the two heterotic theories, have ${\mathcal N}
= 1$ supersymmetry in the ten-dimensional sense.  Since the
minimal 10d spinor is simultaneously Majorana and Weyl, this
corresponds to 16 conserved supercharges.  The other two theories,
called {type IIA} and {type IIB}, have ${\mathcal N} = 2$
supersymmetry (32 supercharges). In the IIA case the two spinors
have opposite handedness so that the spectrum is left-right
symmetric (nonchiral).  In the IIB case the two spinors have the
same handedness and the spectrum is chiral.

In each of these five superstring theories it became clear,
and was largely proved, that
there are consistent perturbation expansions of on-shell scattering amplitudes.
In four of the five cases (heterotic and type II) the fundamental strings are
oriented and unbreakable.  As a result, these theories have particularly simple
perturbation expansions.  Specifically, there is a unique Feynman diagram at
each order of the loop expansion.  The Feynman diagrams
depict string world sheets, and
therefore they are two-dimensional surfaces.  For these four theories the
unique $L$-loop diagram is a closed orientable
genus-$L$ Riemann surface, which can be visualized
as a sphere with $L$ handles.  External (incoming or outgoing) particles are
represented by $N$ points (or ``punctures'') on the Riemann surface.  A
given diagram represents a well-defined integral of dimension $6L + 2N - 6$.  This
integral has no ultraviolet divergences, even though the spectrum contains
states of arbitrarily high spin (including a massless graviton).  From the
viewpoint of point-particle contributions, string and supersymmetry properties
are responsible for incredible cancellations.  Type I superstrings are
unoriented and breakable.  As a result, the perturbation expansion is more
complicated for this theory, and various world-sheet diagrams at a given order
have to be
combined properly to cancel divergences and anomalies .

An important discovery that was made between the two superstring revolutions
is {\em T duality}.  As we explained earlier,
this duality relates two string theories when one spatial dimension forms a
circle (denoted $S^1$).  Then the ten-dimensional geometry is $R^9 \times S^1$.
T duality identifies this string compactification with one of a second string
theory also on $R^9 \times S^1$.  If the radii of the circles in the
two cases are denoted $R_1$ and $R_2$, then
\begin{equation}
R_1 R_2 = \alpha'. \label{Tdual}
\end{equation}
Here $\alpha' = \ell_s^2$ is the universal Regge slope parameter, and $\ell_s$
is the fundamental string length scale (for both string theories).
Note that T duality implies that shrinking the circle to zero in one theory
corresponds to decompactification of the dual theory.

The type IIA and IIB theories are T dual, so compactifying the
nonchiral IIA theory on a circle of radius $R$ and letting $R
\rightarrow 0$ gives the chiral IIB theory in ten dimensions! This
means, in particular, that they should not be regarded as distinct
theories.  The radius $R$ is actually the vacuum value of a scalar
field, which arises as an internal component of the 10d metric
tensor.  Thus the type IIA and type IIB theories in 10d are two
limiting points in a continuous moduli space of quantum vacua. The
two heterotic theories are also T dual, though there are
additional technical details in this case. T duality applied to
the type I theory gives a dual description, which is sometimes
called type I${}^{\prime}$ or IA.

\subsection{M Theory}

In the 1970s and 1980s various supersymmetry and supergravity
theories were constructed. In particular, supersymmetry
representation theory showed that the largest possible spacetime
dimension for a supergravity theory (with spins $\leq 2$) is
eleven.  Eleven-dimensional supergravity, which has 32 conserved
supercharges, was constructed in 1978 by Cremmer, Julia, and
Scherk \cite{Cremmer}. It has three kinds of fields---the graviton
field (with 44 polarizations), the gravitino field (with 128
polarizations), and a three-index gauge field $C_{\mu\nu\rho}$
(with 84 polarizations).  These massless particles are referred to
collectively as the {\em supergraviton}. 11d supergravity is
nonrenormalizable, and thus it cannot be a fundamental theory.
However, we now believe that it is a low-energy effective
description of M theory, which is a well-defined quantum theory.
This means, in particular, that higher-dimension terms in the
effective action for the supergravity fields have uniquely
determined coefficients within the M theory setting, even though
they are formally infinite (and hence undetermined) within the
supergravity context.

Intriguing connections between type IIA string theory and 11d
supergravity have been known for a long time, but the precise
relationship was only explained in 1995. The field equations of
11d supergravity admit a solution that describes a supermembrane.
In other words, this solution has the property that the energy
density is concentrated on a two-dimensional surface.  A 3d
world-volume description of the dynamics of this supermembrane,
quite analogous to the 2d world volume actions of superstrings (in
the GS formalism \cite{Greenc}), was constructed by Bergshoeff,
Sezgin, and Townsend in 1987 \cite{Bergshoeff}. The authors
suggested that a consistent 11d quantum theory might be defined in
terms of this membrane, in analogy to string theories in ten
dimensions. (Most experts now believe that M theory cannot be
defined as a supermembrane theory.)  Another striking result was
that a suitable dimensional reduction of this supermembrane gives
the (previously known) type IIA superstring world-volume action.
For many years these facts remained unexplained curiosities until
they were reconsidered by Townsend \cite{Townsend} and by Witten
\cite{Wittena}. The conclusion is that type IIA superstring theory
really does have a circular 11th dimension in addition to the
previously known ten spacetime dimensions.  This fact was not
recognized earlier because the appearance of the 11th dimension is
a nonperturbative phenomenon, not visible in perturbation theory.

To explain the relation between M theory and type IIA string theory, a good
approach is to identify the parameters that characterize each of them and to
explain how they are related.  Eleven-dimensional supergravity (and hence M
theory, too) has no dimensionless parameters.  The only
parameter  is the 11d Newton constant, which raised to a suitable power
($-1/9$), gives the 11d Planck mass $m_p$.
When M theory is compactified on a
circle (so that the spacetime geometry is $R^{10} \times S^1$) another
parameter is the radius $R$ of the circle.
Now consider the parameters of type IIA superstring theory.  They are the
string mass scale $m_s$, introduced earlier, and the dimensionless string
coupling constant $g_s$.

We can identify compactified M theory with type IIA superstring theory by
making the following correspondences:
\begin{equation}\label{M1}
m_s^2 = 2\pi R m_p^3
\end{equation}
\begin{equation}\label{M2}
g_s = 2\pi Rm_s.
\end{equation}
Using these one can derive
$g_s = (2\pi Rm_p)^{3/2}$ and $
m_s = g_s^{1/3} m_p$.
The latter implies that the 11d Planck length is shorter than the string length
scale at weak coupling by a factor of $(g_s)^{1/3}$.

Conventional string perturbation theory is an expansion in powers of $g_s$ at
fixed $m_s$.  Equation~(\ref{M2}) shows that this is equivalent to an expansion
about $R=0$.  In particular, the strong coupling limit of type IIA superstring
theory corresponds to decompactification of the eleventh dimension,
so in a sense M theory
is type IIA string theory at infinite coupling. (The $E_8 \times E_8$ heterotic
string theory is also eleven-dimensional at strong coupling.)
This explains why the eleventh dimension was not discovered in
studies of string perturbation theory.

These relations encode some interesting facts.  For one thing, the fundamental
IIA string actually {\em is} an M2-brane of M theory with one of its dimensions
wrapped around the circular spatial dimension.
Denoting the string and membrane tensions (energy
per unit volume) by $T_{F1}$ and $T_{M2}$, one deduces that
\begin{equation}
T_{F1} = 2\pi R \, T_{M2}.
\end{equation}
However, $T_{F1} = 2\pi m_s^2$ and $T_{M2} = 2\pi m_p^3$.  Combining these
relations gives eq.~(\ref{M1}).

\subsection{Type II $p$-branes}

Type II superstring theories contain a variety of $p$-brane solutions that preserve
half of the 32 supersymmetries. These are solutions in which
the energy  is concentrated on a $p$-dimensional spatial
hypersurface. (The  world volume has $p+1$ dimensions.)
The corresponding solutions of  supergravity
theories were constructed in 1991 by Horowitz and Strominger \cite{Horowitz}.
A large class of these $p$-brane excitations are called
{\em D-branes} (or D$p$-branes when we want to specify the dimension),
 whose tensions are given by
\begin{equation}
T_{Dp} = 2\pi {m_s^{p+1}}/{g_s}.
\end{equation}
This dependence on the coupling constant is one of the characteristic features
of a D-brane.  Another characteristic feature of D-branes
is that they carry a charge that couples to a gauge field
in the RR sector of the theory \cite{Polchinski}.
The particular RR gauge fields that
occur imply that $p$ takes even values in the IIA theory
and odd values in the IIB theory.

In particular, the D2-brane of the type IIA theory
corresponds to the supermembrane of M theory, but now
in a background geometry in which one of the transverse dimensions is a circle.
The tensions check, because (using eqs.~(\ref{M1}) and~(\ref{M2}))
\begin{equation}
T_{D2} = 2\pi {m_s^3}/{g_s} = 2\pi m_p^3 = T_{M2}.
\end{equation}
The mass of the first Kaluza--Klein excitation
of the 11d supergraviton is $1/R$.  Using eq.~(\ref{M2}),
we see that this can be identified with the D0-brane.
More identifications of this type arise when we consider the magnetic dual of
the M theory supermembrane, which is a five-brane, called the
M5-brane.\footnote{In general, the magnetic dual of a $p$-brane in $d$
dimensions is a $(d - p - 4)$-brane.}  Its tension is $T_{M5} = 2\pi m_p^6$.
Wrapping one of its dimensions around the circle gives the D4-brane, with
tension
\begin{equation}
T_{D4} = 2\pi R \,T_{M5} = 2\pi m_s^5/g_s.
\end{equation}
If, on the other hand, the M5-frame is not wrapped around the circle, one
obtains the NS5-brane of the IIA theory with tension
\begin{equation}
T_{NS5} = T_{M5} = 2\pi m_s^6/g_s^2.
\end{equation}

To summarize, type IIA superstring theory is M theory compactified on a circle
of radius $R=g_s \ell_s$.
 M theory is believed to be a well-defined quantum theory in 11d, which is
approximated at low energy by 11d supergravity.  Its excitations are the
massless supergraviton, the M2-brane, and the M5-brane.  These account both for
the (perturbative) fundamental string of the IIA theory and for many of its
nonperturbative excitations.  The identities that we have presented here are exact,
because they are protected by supersymmetry.

\subsection{Type IIB Superstring Theory}

Type IIB superstring theory, which is the other maximally supersymmetric
string theory with
32 conserved supercharges, is also 10-dimensional, but unlike the IIA
theory its two supercharges have the same handedness.  At low-energy,
type IIB superstring theory is approximated by type IIB supergravity,
just as 11d supergravity approximates M theory.  In each case the
supergravity theory is only well-defined as a classical field theory, but still
it can teach us a lot.  For example, it can be used to construct $p$-brane
solutions and compute their tensions.  Even though such solutions are
only approximate, supersymmetry considerations ensure that the tensions,
which are related to the kinds of conserved charges the $p$-branes carry, are exact.
Since the IIB spectrum
contains massless chiral fields, one should check whether there are anomalies
that break the gauge invariances---general coordinate invariance, local Lorentz
invariance, and local supersymmetry.  In fact, the UV finiteness of the
string theory Feynman diagrams ensures that all anomalies must
cancel, as was verified from a field theory viewpoint by
Alvarez-Gaum\'e and Witten \cite{Alvarez}.

Type IIB superstring theory or supergravity contains two scalar fields, the
dilation $\phi$ and an {axion} $\chi$, which are conveniently combined in a
complex field
\begin{equation}
\rho = \chi + ie^{-\phi}.
\end{equation}
The supergravity approximation has an $SL(2,R)$ symmetry
that transforms this field nonlinearly:
\begin{equation}
\rho \rightarrow \frac{a\rho + b}{c\rho + d},
\end{equation}
 where $a,b,c,d$ are real numbers satisfying $ad - b c = 1$.  However, in the
quantum string theory this symmetry is broken to the discrete subgroup
$SL(2,Z)$ \cite{Hull}, which means that $a,b,c,d$ are
restricted to be integers.  Defining
the vacuum value of the $\rho$ field to be
\begin{equation}
\langle \rho \rangle = \frac{\theta}{2\pi} + \frac{i}{g_s},
\end{equation}
the $SL(2,Z)$ symmetry transformation $\rho \rightarrow \rho + 1$
implies that $\theta$
is an angular coordinate.  Moreover, in the special case $\theta =
0$, the symmetry transformation $\rho \rightarrow - 1/\rho$ takes $g_s
\rightarrow 1/g_s$.  This symmetry, called {\em S duality}, implies that
coupling constant $g_s$ is equivalent to coupling constant $1/g_s$, so
that, in the case of Type II superstring theory,
the weak coupling expansion and the strong coupling expansion are
identical! (An analogous S-duality transformation relates the Type I
superstring theory to the $SO(32)$ heterotic string theory.)

Recall that the type IIA and type IIB
superstring theories are T dual, meaning that if they are compactified on
circles of radii $R_A$ and $R_B$ one obtains equivalent theories for the
identification $R_AR_B = \ell_s^2$.  Moreover, we saw that the
type IIA theory is actually M theory compactified on a circle.  The latter fact
encodes nonperturbative information.  It turns out to be very useful to combine
these two facts and to consider the duality between M theory compactified on a
torus $(R^9 \times T^2)$ and type IIB superstring theory compactified on a
circle $(R^9 \times S^1)$.

A torus can be described as the complex plane modded out by the
equivalence relations $z \sim z + w_1$ and $z \sim z + w_2$.  Up
to conformal equivalence, the periods $w_1$ and $w_2$ can be
replaced by $1$ and $\tau$, with Im $\tau > 0$.  In this
characterization $\tau$ and $\tau' = (a\tau + b)/(c\tau + d)$,
where $a,b,c,d$ are integers satisfying $ad - b c = 1$, describe
equivalent tori.  Thus a torus is characterized by a modular
parameter $\tau$ and an $SL(2,Z)$ modular group.  The natural, and
correct, conjecture at this point is that one should identify the
modular parameter $\tau$ of the M theory torus with the parameter
$\rho$ that characterizes the type IIB vacuum \cite{Schwarza, Aspinwall}.
Then the duality
of M theory and type IIB superstring theory gives a geometrical
explanation of the nonperturbative S duality symmetry of the IIB
theory:  the transformation $\rho \rightarrow - 1/\rho$, which
sends $g_s \rightarrow 1/g_s$ in the IIB theory, corresponds to
interchanging the two cycles of the torus in the M theory
description.  To complete the story, we should relate the area of
the M theory torus $(A_M)$ to the radius of the IIB theory circle
$(R_B)$.  This is a simple consequence of formulas given above
\begin{equation}
m_p^3 A_M = (2 \pi R_B)^{-1}.
\end{equation}
Thus the limit $R_B \rightarrow 0$, at fixed $\rho$, corresponds to
decompactification of the M theory torus, while preserving its shape.
Conversely, the limit $A_M \rightarrow 0$ corresponds to decompactification of
the IIB theory circle.
The duality can be explored further by matching the various $p$-branes in 9
dimensions that can be obtained from either the M theory or the IIB theory
viewpoints.
When this is done, one finds that everything matches nicely and
that one deduces various relations among tensions \cite{Schwarzb}.

Another interesting fact about the IIB theory is that it contains
an infinite family of strings labeled by a pair of  integers
$(p,q)$ with no common divisor \cite{Schwarza}. The $(1,0)$ string can be
identified as the fundamental IIB string, while the $(0,1)$ string
is the D-string.  From this viewpoint, a $(p,q)$ string can be
regarded as a bound state of $p$ fundamental strings and $q$
D-strings \cite{Wittenb}. These strings have a very simple interpretation in the
dual M theory description.  They correspond to an M2-brane with
one of its cycles wrapped around a $(p,q)$ cycle of the torus. The
minimal length of such a cycle is proportional to $|p+q \tau|$,
and thus (using $\tau = \rho$) one finds that the tension of a
$(p,q)$ string is given by
\begin{equation}
T_{p,q} = 2\pi|p + q\rho| m_s^2. \label{pqtension}
\end{equation}

Imagine that you lived in the 9-dimensional world that is described
equivalently as M theory compactified on a torus or as the type IIB
superstring theory
compactified on a circle.  Suppose, moreover, you had very high energy
accelerators with which you were going to determine the ``true'' dimension of
spacetime.  Would you conclude that 10 or 11 is the correct answer?  If either
$A_M$ or $R_B$ was very large in Planck units there would be a natural choice,
of course.  But how could you decide otherwise?  The answer is that either
viewpoint is equally valid.  What determines which choice you make is which of
the massless fields you regard as ``internal'' components of the metric tensor
and which ones you regards as matter fields.
Fields that are metric components in one
description correspond to matter fields in the dual one.

\subsection{The D3-Brane and ${\mathcal N}=4$ Gauge Theory}

D-branes have a number of special properties, which make them especially
interesting.   By definition, they are branes on which strings can end---D
stands for {\em Dirichlet} boundary conditions.  The end of a string carries a
charge, and the D-brane world-volume theory contains
a $U(1)$ gauge field that carries the
associated flux.  When $n$ D$p$-branes are coincident, or parallel and nearly
coincident, the associated $(p + 1)$-dimensional world-volume theory is a
$U(n)$ gauge theory \cite{Wittenb}.  The $n^2$ gauge bosons $A_\mu^{ij}$ and their
supersymmetry partners arise as the ground states of oriented strings running
from the $i$th D$p$-brane to the $j$th D$p$-brane.  The diagonal elements,
belonging to the Cartan subalgebra, are massless.  The field
$A_\mu^{ij}$ with $i \not= j$ has a mass proportional to the separation of the
$i$th and $j$th branes.

The $U(n)$ gauge theory associated with a stack of $n$ D$p$-branes has maximal
supersymmetry (16 supercharges).  The low-energy effective theory, when the
brane separations are small compared to the string scale, is supersymmetric
Yang--Mills theory.  These theories can be constructed by dimensional reduction
of 10d supersymmetric $U(n)$ gauge theory to $p+1$ dimensions.
A case of particular interest, which we shall now focus on, is $p = 3$.  A stack of
$n$ D3-branes in type IIB superstring theory
has a decoupled ${\mathcal N} = 4, $ $d = 4$
$U(n)$ gauge theory associated to it.  This gauge theory has a number of special
features.  For one thing, due to boson--fermion cancellations, there are no
$UV$ divergences at any order of perturbation theory.  The beta function
$\beta(g)$ is identically zero, which implies that the
theory is scale invariant.  In
fact, ${\mathcal N}=4, $ $d=4$ gauge theories are
conformally invariant.  The conformal
invariance combines with the supersymmetry to give a superconformal symmetry,
which contains 32 fermionic generators.
Another important property of ${\mathcal N}=4$, $ d=4$ gauge theories is an
electric-magnetic duality, which extends to an $SL(2,Z)$ group of dualities.
Now consider the ${\mathcal N}=4 $ $U(n)$ gauge theory associated to a stack of $n$
D3-branes in type IIB superstring theory.
There is an obvious identification, that turns out to be correct.
Namely, the $SL(2,Z)$ duality of the gauge theory is induced from that of the
ambient type IIB superstring theory.
The D3-branes themselves are invariant under $SL(2,Z)$ transformations.

As we have said, a fundamental $(1,0)$ string can end on a D3-brane.  But by
applying a suitable $SL(2,Z)$ transformation, this configuration is transformed
to one in which a $(p,q)$ string ends on
the D3-brane.  The charge on the end of this string describes a dyon with
electric charge $p$ and magnetic charge $q$, with respect to the
appropriate gauge field.
More generally, for a stack of $n$
D3-branes, any pair can be connected by a $(p,q)$ string.  The mass is
proportional to the length of the string times its tension, which we saw is
proportional to $|p + q\rho|$.  In this way one sees that the electrically
charged particles, described by fundamental fields, belong to infinite
$SL(2,Z)$ multiplets.  The other states are nonperturbative excitations of the
gauge theory.  The field configurations that describe them
preserve half of the supersymmetry.  As a
result their masses are given exactly by the
considerations described above.
An interesting question, whose answer was unknown until recently, is whether
${\mathcal N}=4 $  gauge theories in four dimensions
also admit nonperturbative excitations that preserve
1/4 of the supersymmetry.  The answer turns out to be that they do, but only
if $n\geq 3$. This result has a nice dual description in terms of three-string
junctions \cite{Bergman}.

\subsection{Conclusion}

In this lecture we have described some of the interesting advances
in understanding superstring theory that have taken place in the past
few years. The emphasis has been on the nonperturbative appearance
of an eleventh dimension in type IIA superstring theory, as well as
its implications when combined with superstring T dualities. In particular,
we argued that there should be a consistent quantum vacuum, whose
low-energy effective description is given by 11d supergravity.

What we have described makes a convincing self-consistent picture, but it
does not constitute a complete formulation of M theory. In the past several years
there have been some major advances in that direction, which we will
briefly mention here. The first, which goes by the name
of {\em Matrix Theory},
bases a formulation of M theory in flat 11d spacetime in terms of the
supersymmetric quantum mechanics of $N$ D0-branes in the large $N$ limit.
Matrix Theory has passed
all tests that have been carried out, some of which are very nontrivial.
The construction has a nice
generalization to describe compactification of M theory
on a torus $T^n$.
However, it does not seem to be useful for $n > 5$, and other
compactification manifolds are (at best) awkward to handle. Another
shortcoming of this approach is that it treats the eleventh dimension
differently from the other ones.

Another proposal relating superstring and M theory backgrounds to large $N$
limits of certain field theories has been put forward
by Maldacena in 1997 \cite{JM} and made
more precise by Gubser, Klebanov, and Polyakov \cite{GKP}, and
by Witten \cite{W1} in 1998.
(For a review of this subject, see \cite{MAGOO}.)
In this approach, there is a conjectured duality ({\it i.e.}, equivalence) between a
conformally invariant field theory (CFT) in $d$ dimensions and
type IIB superstring theory
or M theory on an Anti-de-Sitter space (AdS) in $d+1$ dimensions. The remaining
$9-d$ or $10-d$ dimensions form a compact space, the simplest cases being spheres.
Three examples with unbroken supersymmetry are $AdS_5 \times S^5$,
$AdS_4 \times S^7$, and $AdS_7 \times S^4$.
This approach is sometimes referred to as {\em AdS/CFT duality}.
This is an extremely active and very promising subject. It has already taught us
a great deal about the large $N$ behavior of various gauge theories.
As usual, the easiest theories to study are ones with a lot of supersymmetry,
but it appears that in this approach supersymmetry breaking is more accessible than
in previous ones. For example, it might someday be possible to construct the QCD
string in terms of a dual AdS gravity theory, and use it
to carry out numerical calculations of the hadron spectrum.
Indeed, there have already been some
preliminary steps in this direction.

Despite all of the successes that have been achieved
in advancing our understanding of
superstring theory and M theory, there clearly is still a long way
to go.  In particular, despite much effort and several imaginative proposals,
we still do not have a convincing mechanism for ensuring the vanishing
(or extreme smallness) of the cosmological constant for nonsupersymmetric vacua.
Superstring theory is a field with very ambitious goals. The remarkable fact is
that they still seem to be realistic. However,
it may take a few more revolutions before they are attained.

\end{document}